\documentclass[showpacs,amsmath,amssymb,twocolumn,superscriptaddress,notitlepage,preprintnumbers,longbibliography]{revtex4-1}

\usepackage{braket}
\usepackage{physics}
\usepackage{mathtools}
\usepackage{diagbox}
\usepackage[utf8]{inputenc}
\usepackage{algorithm}
\usepackage{color}
\usepackage{amssymb}
\usepackage{amsmath}%
\usepackage{algpseudocode}
\usepackage{algorithm}
\usepackage{qcircuit}
\usepackage{orcidlink}
\usepackage{titlesec}
\usepackage{cprotect}

\newcommand{\ud}{\mathrm{d}}
\newcommand{\DD}{\mathcal{D}}

\newcommand{\OO}{\mathcal{O}}

\newcommand{\bos}{\boldsymbol}
\newcommand{\shortxymatrix}[1]{\xymatrix@1@C=.6cm{#1}}
\newcommand{\EZero}{\shortxymatrix{\ar@{-}[r]&}}

\newcommand{\Var}{\mathrm{Var}}

\newcommand{\ii}{\mathbf{i}}

\newcommand{\sX}{\hat{\sigma}^x}
\newcommand{\sY}{\hat{\sigma}^y}
\newcommand{\sZ}{\hat{\sigma}^z}

\usepackage{hyperref}
\hypersetup{colorlinks=true, linkcolor=blue, citecolor=blue, urlcolor=magenta }

\newcommand{\PKU}{Center on Frontiers of Computing Studies, Peking University, Beijing 100871, China}

\newcommand{\PKUCS}{School of Computer Science, Peking University, Beijing 100871, China}

\usepackage[author=cxx,color=blue]{changes}

\begin{document}

\title{Computing $n$-time correlation functions without ancilla qubits}

\author{Xiaoyang Wang\orcidlink{0000-0002-2667-1879} }

\affiliation{RIKEN Center for Interdisciplinary Theoretical and Mathematical Sciences (iTHEMS), Wako 351-0198, Japan}
\affiliation{RIKEN Center for Computational Science (R-CCS), Kobe 650-0047, Japan}

\author{Long Xiong}

\affiliation{\PKU}
\affiliation{\PKUCS}

\author{Xiaoxia Cai}
\thanks{xxcai@ihep.ac.cn}
\affiliation{Institute of High Energy Physics, Chinese Academy of Sciences, Beijing 100049, China}

\author{Xiao Yuan}
\thanks{xiaoyuan@pku.edu.cn}
\affiliation{\PKU}
\affiliation{\PKUCS}

\date{\today}
\begin{abstract}
The $n$-time correlation function is pivotal for establishing connections between theoretical predictions and experimental observations of a quantum system. Conventional methods for computing $n$-time correlation functions on quantum computers, such as the Hadamard test, generally require an ancilla qubit that controls the entire system~---~an approach that poses challenges for digital quantum devices with limited qubit connectivity, as well as for analog quantum platforms lacking controlled operations. Here, we introduce a method to compute $n$-time correlation functions using only unitary evolutions on the system of interest, thereby eliminating the need for ancillas and the control operations. This approach substantially relaxes hardware connectivity requirements for digital processors and enables more practical measurements of $n$-time correlation functions on analog platforms. We demonstrate our protocol on IBM quantum hardware up to 12 qubits to measure the single-particle spectrum of the Schwinger model and the out-of-time-order correlator in the transverse-field Ising model. In the demonstration, we further introduce an error mitigation procedure based on signal processing that integrates signal filtering and correlation analysis, and successfully reproduces the noiseless simulation results from the noisy hardware. Our work highlights a route to exploring complex quantum many-body correlation functions in practice, even in the presence of realistic hardware limitations and noise.
\end{abstract}

\maketitle

Correlation functions are essential for characterizing quantum dynamics and are widely used in many-body physics. Practically, their values can be directly compared to experimental data, such as in spectroscopy~\cite{mukamel_2015,mchale_2017}. Particularly, two-time correlation functions are key to extracting information on conductivity and magnetization via the Kubo relation~\cite{doi:10.1143/JPSJ.12.570}. Additionally, dynamical structure factors, derived from density-density or spin-spin correlations, provide vital insights into material properties~\cite{Baez_2020,Boothroyd_2020,Chiesa_2019}.
Moreover, higher-order correlation functions, such as the out-of-time-order correlator (OTOC), have gained attention for their ability to probe information scrambling in condensed matter systems~\cite{Swingle_2016,Swingle_2018,Swingle_2018b,PhysRevResearch.7.023032}. In particle physics, time correlation functions are similarly crucial, relating to particle scattering amplitudes via the Lehmann–Symanzik–Zimmermann reduction formula~\cite{Peskin:1995ev}.

Calculating $n$-time correlation functions requires simulating the Schr\"odinger evolution of quantum systems, a task that is challenging for classical computers. This makes such simulations a promising avenue for demonstrating practical quantum advantages on currently available hardware~\cite{quantum_ring2021, Daley_22, IBM_Eagle23, doi:10.1126/science.ado6285, JinZhaoSun2025, shinjo2024unveilingcleantwodimensionaldiscrete,Efekan_2024,yoshioka2024hunting}.
Several studies have explored quantum circuit-based methods~\cite{Roggero_2019,Kosugi_2020,Ciavarella_2020, Chen_2021,sunjinzhao_2025} and weak measurements~\cite{PfenderMatthias_2019,WangPing_2019,MeinelJonas_2022,PhysRevLett.132.200802,https://doi.org/10.1002/qute.202300286} for computing time correlation functions. A commonly used approach is the Hadamard test~\cite{PhysRevA.65.042323, PhysRevLett.113.020505,PhysRevResearch.2.033281}, which, however, generally requires control operations between an ancillary qubit and the target system. When correlation function operators involve non-local properties, such as momentum-dependent observables, long-range control gates are needed. These gates present significant implementation challenges for practical quantum devices, which primarily rely on neighboring qubit interactions~\cite{childs_et_al:LIPIcs.TQC.2019.3, Bapat2021quantumroutingfast, grigoryan2022algorithmcircuitnestingdoubled, e106-a_2_124}. Moreover, the need for controlled operations in the Hadamard test limits its applicability on analog quantum platforms~\cite{PhysRevLett.129.160602, Liang2025, 10.5555/3600270.3600610, 10.1063/5.0179540, Paula_2024,xukai2023a}. Recent work~\cite{Lorenzo_2024} also points out that the use of auxiliary qubits in such measurements restricts the ability to capture long-time dynamics, further limiting the approach’s effectiveness in certain quantum simulations.

\begin{figure*}
    \centering
    \includegraphics[width=0.98\textwidth]{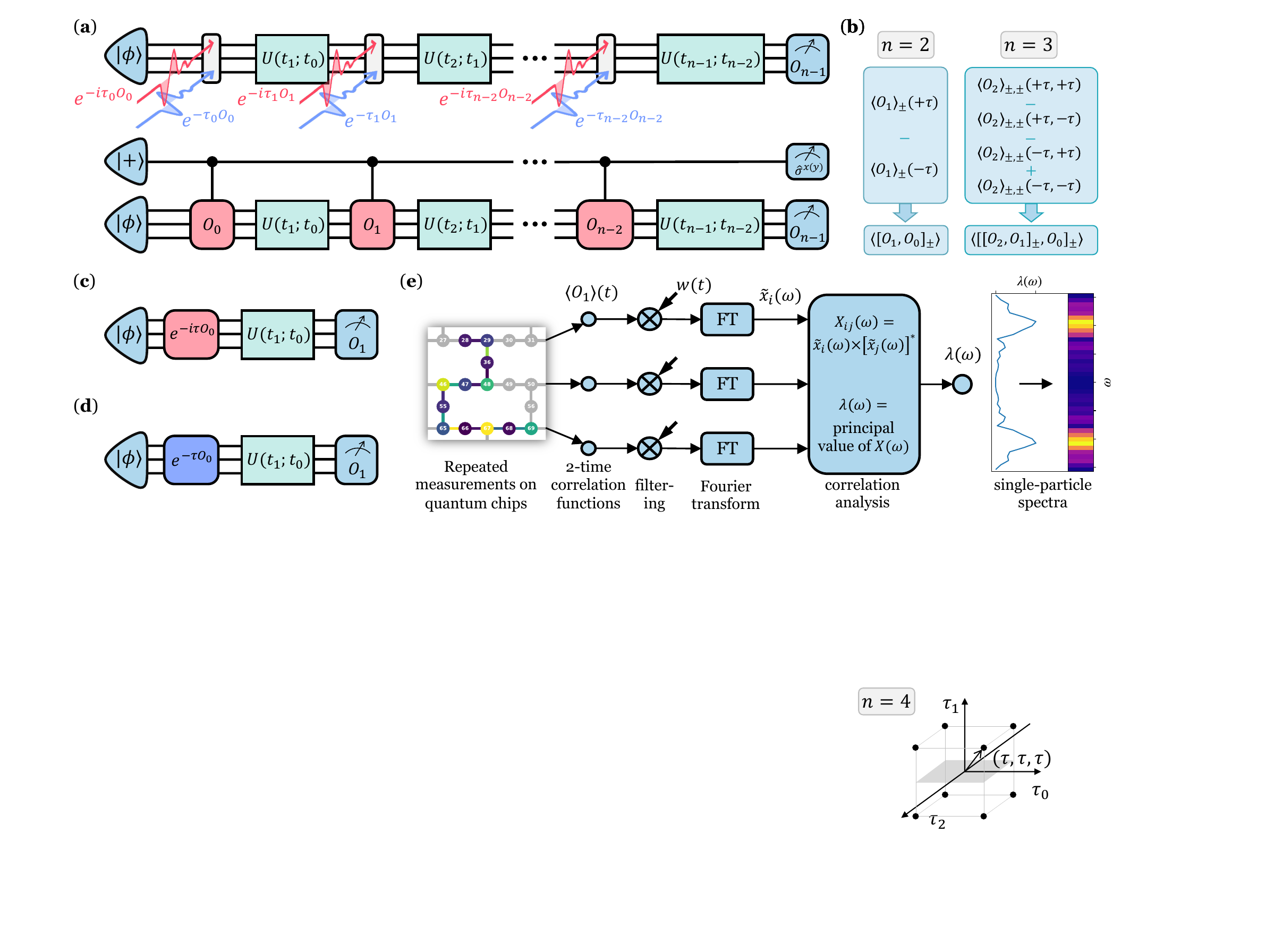}
    \caption{Schematic of the method. (\textbf{a}) The ancilla-free circuit to measure the $n$-time correlation function (upper panel) and the standard Hadamard test circuit using an ancilla qubit (lower panel). Our approach replaces the control operations in the Hadamard test with the real- or imaginary-time evolution of the operator $O_j$, realized using the red vertical pulses or the blue horizontal pulses, respectively. (\textbf{b}) Computing nested $n$-time commutators and anti-commutators using the parameter-shift rule. $\tau$ is an arbitrary real number of $\OO(1)$. Quantum circuits to measure (\textbf{c}) $2$-time commutators and (\textbf{d}) $2$-time anti-commutators. (\textbf{e}) A classical signal-processing method implemented to improve the single-particle spectra by combining signal filtering and correlation analysis. The spectra are derived from the Fourier transformation of the $2$-time correlation functions repeatedly measured on quantum chips.}
    \label{fig:main-circuit}
\end{figure*}


Here, we address these limitations by replacing the long-range control operations of the Hadamard test with real-time and quantum imaginary-time evolution (QITE)~\cite{McArdle_19, Motta_20} only on the target system (Fig.~\ref{fig:main-circuit}(\textbf{a})). 
This is based on a crucial insight of the intimate correspondence between (real and imaginary) time
evolutions and (normal and anti-) commutators.
We further introduce a novel parameter-shift rule and prove that $n$-time correlation functions can be represented through linear combinations of observable expectations (Fig.~\ref{fig:main-circuit}(\textbf{b})).
We demonstrate the effectiveness of our method on IBM quantum hardware with systems up to 12 qubits, calculating the single-particle spectra of the Schwinger model and measuring both the real and imaginary components of the out-of-time-order correlator in the transverse-field Ising model. We introduce error suppression techniques, which filters individual signals and analyzes correlations across multiple signals (Fig.~\ref{fig:main-circuit}(\textbf{e})), significantly reducing hardware noise. Our results validate the feasibility of this approach for practical quantum devices and its potential for digital-analog quantum computing platforms~\cite{PhysRevResearch.6.013280, Kumar_2025, PRXQuantum.2.020328, PhysRevA.101.022305}.



\vspace{0.2cm}

\noindent \textbf{\noindent Framework~---}
Here, the 2-time correlation function is considered as an example; the general case is deferred to Appendix A of the End Note and Supplemental Material~\cite{supp}. The key idea is to first represent correlation functions as a linear combination of nested commutators and anti-commutators. For each commutator and anti-commutator, it is shown that each equals the partial derivative with respect to real-time and imaginary-time evolutions, respectively. These derivatives can, in turn, be expressed as linear combinations of real- and imaginary-time evolutions via a novel parameter-shift rule. Consequently, correlation functions are obtained from linear combinations of measurements under real- and imaginary-time evolutions in Fig.~\ref{fig:main-circuit}(\textbf{a}).

Specifically, consider the 2-time correlation function $\bra{\phi} O_{1}(t_{1}) O_{0}(t_{0})\ket{\phi}$, where $O_{j}(t_{j}) = U^{\dagger}(t_j;t_0)O_{j}U(t_j;t_0)$ is the operator in the Heisenberg picture, $U(t_i; t_j)$ is the unitary evolution of the system from $t_j$ to $t_i$, and $\ket{\phi}$ is the input state.
Based on the definition of commutators and anti-commutators, we have $O_1(t_1)O_0(t_0) =\frac{1}{2}([O_1(t_1),O_0(t_0)]_-+[O_1(t_1),O_0(t_0)]_+)$. Therefore, we only need to focus on how to measure the commutator $\bra{\phi}[O_1(t_1),O_0(t_0)]_-\ket{\phi}$ and the anti-commutator $\bra{\phi}[O_1(t_1),O_0(t_0)]_+\ket{\phi}$.


The key insight draws from the following relation: commutator and anti-commutator can be derived from the partial derivatives with respect to the real and imaginary time evolutions, i.e., 
\begin{equation}
    \begin{aligned}
    [O_1(t_1), O_0]_- &=i\partial_\tau (e^{i\tau O_0}O_1(t_1) e^{-i\tau O_0})|_{\tau=0};\\
    [O_1(t_1), O_0]_+  &= -\partial_\tau (e^{-\tau O_0}O_1(t_1) e^{-\tau O_0})|_{\tau=0}.
    \label{eq:partial-derivative-2-time}
\end{aligned}
\end{equation}
Then, the 2-time correlation function can be obtained from the partial derivatives of real and imaginary time evolutions $\bra{\phi}e^{i\tau O_0}O_1(t_1) e^{-i\tau O_0}\ket{\phi}$ and $\bra{\phi}e^{-\tau O_0}O_1(t_1) e^{-\tau O_0}\ket{\phi}$.

To approximate the partial derivatives, the naive finite-difference approach may introduce large statistical and truncation errors. Instead, without loss of generality, we can assume $O_0$ to be a Pauli string, and the partial derivatives can be obtained accurately based on a novel parameter-shift rule~\cite{supp}, originally developed for variational quantum algorithms~\cite{PhysRevLett.118.150503, crooks2019gradientsparameterizedquantumgates}. We further generalize it to evaluate high-order derivatives for general $n$-time correlation functions with respect to both real and imaginary evolution time,
as illustrated in Fig.~\ref{fig:main-circuit}(\textbf{b}) and detailed in Appendix B of the End Note. Finally, we can obtain $2$-time correlation functions via the expectations of $\bra{\phi}e^{i\tau O_0}O_1(t_1) e^{-i\tau O_0}\ket{\phi}$ and $\bra{\phi}e^{-\tau O_0}O_1(t_1) e^{-\tau O_0}\ket{\phi}$, using the ancilla-free circuits in Fig.~\ref{fig:main-circuit}(\textbf{c},\textbf{d}), respectively.

\begin{figure*}
    \centering
    \includegraphics[width=0.98\textwidth]{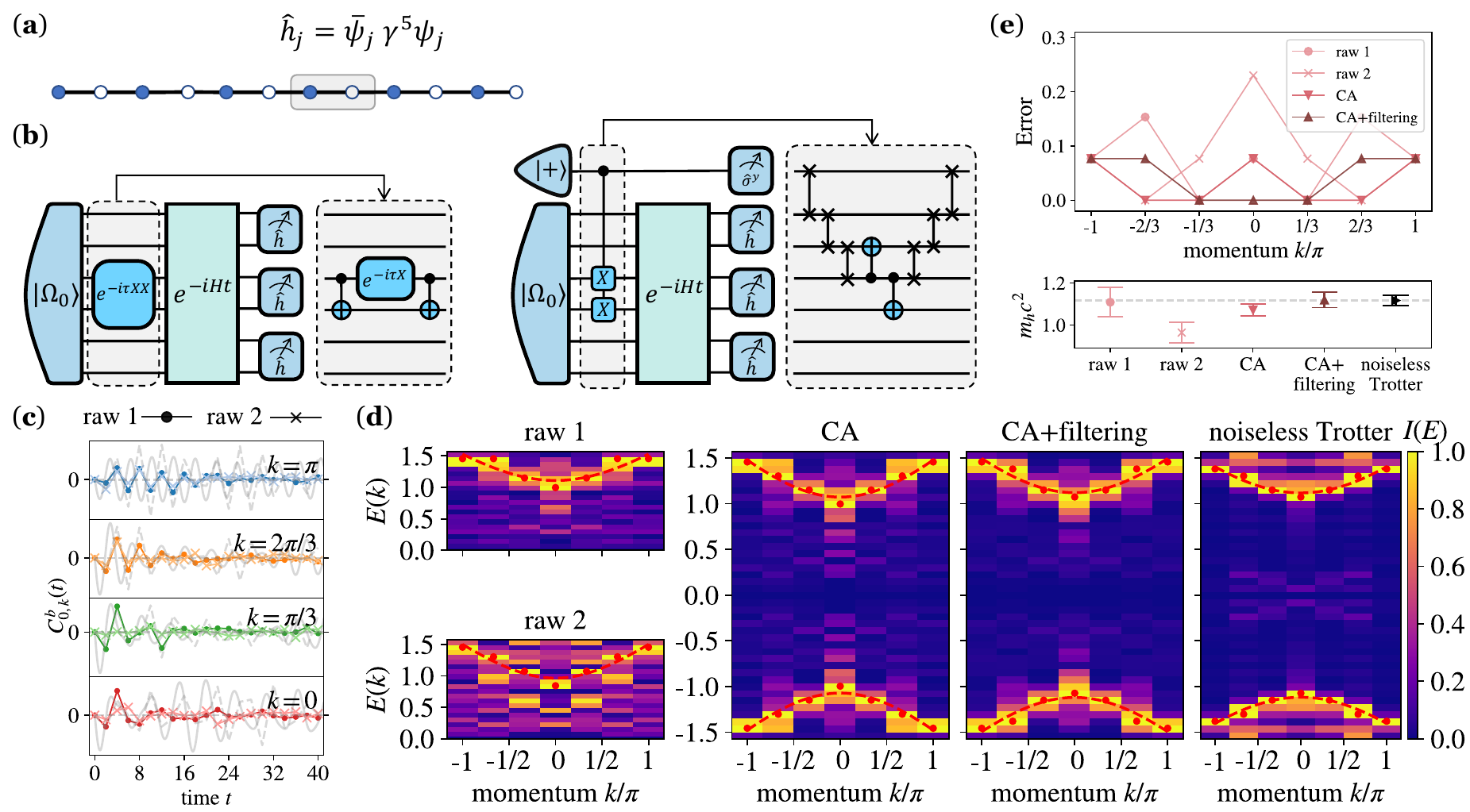}
    \cprotect\caption{Measuring the hadron spectrum of the Schwinger model. (\textbf{a}) Lattice and hadron operator of the Schwinger model. (\textbf{b}) Comparison of the measurement circuits for the hadron spectrum using the ancilla-free circuit (left panel) and Hadamard test (right panel), and their physical realization on digital quantum chips with linear qubit connectivity (dashed boxes). (\textbf{c}) Measurement results of 2-time correlation function $C^b_{k,0}(t)$ on \verb|ibm_torino| versus time $t$ with four different momentum $k$. The grey solid and dashed lines denote exact diagonalization and noiseless Trotter results. We perform the measurement twice for correlation analysis (CA), labeled by \textit{raw 1} and \textit{raw 2}. (\textbf{d}) Normalized hadron spectrum of the two raw results, classically processed using CA, CA and filtering, and the noiseless Trotter spectrum. The highest energy peak of each $k$ is marked by a red dot, and fitted by the red dashed line using the hadron dispersion relation. (\textbf{e}) Effect of the classical signal-processing method. The error (upper panel) is the position of energy peaks between the measured and the noiseless Trotter results. $m_h c^2$ (lower panel) is obtained from the fitted hadron dispersion relation, with the error bars denoting the fitting errors. The grey dashed line denotes the center value of the noiseless Trotter result.}
    \label{fig:Schwinger-result}
\end{figure*}


The real-time evolution in Fig.~\ref{fig:main-circuit}(\textbf{c}) can be realized using standard Hamiltonian simulation algorithms. For imaginary-time evolution in Fig.~\ref{fig:main-circuit}(\textbf{d}), we could exploit the quantum imaginary-time evolution (QITE) algorithm to find a unitary gate $U^{(+)}_{O_0}$ such that
\begin{align}
    U^{(+)}_{O_0}\ket{\phi} \equiv \frac{e^{-\tau O_0}}{\sqrt{\bra{\phi} e^{-2\tau O_0} \ket{\phi}}}\ket{\phi}.
\end{align}
This is possible provided that $O_0$ acts locally on at most $k$ qubits and the initial state $\ket{\phi}$ has a finite correlation length~\cite{Motta_20}. These conditions are satisfied in many physical applications of $2$-time correlation functions, e.g., $\ket{\phi}$ is the ground state of a gapped quantum system. In case the initial state is not finitely correlated, we show that the QITE can also be realized by the mid-circuit measurement without involving ancilla qubits~\cite{supp}. Alternatively, the anti-commutator can be efficiently measured using dissipative techniques~\cite{Pan_2020} with hardware-native dissipation or engineered dissipation~\cite{Zhao2025}.



To demonstrate the scalability of the ancilla-free measurement method, we analyze the errors from three sources: QITE, Trotter decomposition, and a finite number of shots. We show that for physically relevant applications and spatially or momentum localized operators $O_j$, the required number of shots and Trotter steps grow at most polynomially with the system size (see Supplemental Material~\cite{supp} for details). Thus, the method is scalable to large systems. 
By analyzing the QITE gate counts, the practical advantage of the ancilla-free method over the standard Hadamard test method can be guaranteed with two conditions: (1) the quantum chip has a finite qubit connectivity on a $d$-dimensional lattice; (2) QITE's initial state has a finite correlation length $\xi=\OO((\log N/2kd)^{1/d})$, where $N$ is the number of qubits or the system size. The second condition can be circumvented using the mid-circuit measurement for QITE.

Practically, our approach can be applied to measure single-particle spectra of quantum many-body systems via Fourier transforms of 2-time correlation functions. To mitigate noise in spectra obtained from quantum hardware, we introduce a robust signal-processing technique that combines signal filtering with correlation analysis~\cite{4309918, chan2024algorithmicshadowspectroscopy, JinZhaoSun2025}, as shown in Fig.~\ref{fig:main-circuit}(\textbf{e}). This technique involves three steps: (1) repeatedly measuring 2-time correlation functions on quantum hardware; (2) applying a window function $\mathrm{w}(t)$ to enhance short-time signals and suppress long-time noise; and (3) performing correlation analysis on the Fourier-transformed functions to reduce stochastic noise. In our hardware demonstration, this technique significantly improves the noisy single-particle spectra.

Next, we demonstrate the applicability of our method in measuring single-particle spectra and OTOC on the 133-qubit \verb|ibm_torino| IBM quantum hardware~\cite{Qiskit,supp}.


\vspace{0.2cm}

\vspace{0.2cm}
\noindent\textbf{Hadron spectrum ---}
We first apply our method to measure the single-particle spectra of the Schwinger model~\cite{zhaihui2024}, a lattice gauge theory exhibiting fermion confinement and string-breaking effects as observed in quantum chromodynamics~\cite{Mathis_20, PRXQuantum.5.020315,PhysRevD.108.034501, PhysRevD.109.114510,PRXQuantum.5.037001,KuhnStefan2015}. Due to fermion confinement, the Schwinger model’s elementary excitations are bosonic hadrons composed of two fermions. Measuring the hadron spectrum is a key task in applying quantum computation to lattice gauge theories~\cite{Itou2024, Ghim:2024pxe}. We use the truncated Hamiltonian $H_{\text{SW}}^{\text{tr}}$~\cite{PhysRevD.109.114510}, discretized on a one-dimensional lattice with open boundary conditions and $L$ spatial sites, where two qubits are assigned to each site (Fig.~\ref{fig:Schwinger-result}(\textbf{a})). The Hamiltonian and the hadron operator $\hat{h}_j = \bar{\psi}_j \gamma^5 \psi_j$ at site $j$ are mapped to spin operators, as described in Supplemental Material~\cite{supp}.

We use the real-time evolution of the Schwinger model to measure the $2$-time correlation function 
$
    C^b_{0,k}(t)= -i\sum_j e^{-ik(j+j_0)}\bra{\Omega_0}[\hat{h}_j(t),\hat{h}_{j_0}(0)]_-\ket{\Omega_0}
$, where the lattice momentum $k=2\pi n/L,n\in[-L/2,L/2]$, and $\ket{\Omega_0}=\ket{01\cdots 01}$ is the bare vacuum of the Schwinger model. We choose $j_0=\lceil L/2\rceil$ to be at the center of the lattice to minimize the boundary effects. $C^b_{0,k}(t)$ is a good approximation to hadron Green's function, from which the hadron spectrum can be obtained. 


In Fig.~\ref{fig:Schwinger-result}(\textbf{b}), we compare the measurement circuits of $C^b_{0,k}(t)$ using the ancilla-free circuit (left panel) and Hadamard test (right panel). Assume that the digital quantum device has linear qubit connectivity, the long-range control operation of $\hat{h}_{j_0}$ in the right panel should be realized by SWAP gates, whose number increases linearly with the system size $L$, making the measurement results noisy. In contrast, the ancilla-free circuit only requires local evolution that can be realized by two CNOT gates. This improvement is significant for practical quantum computers, for which the two-qubit gates are the primary source of noise.

We measure \( C^b_{0,k}(t) \) on the \texttt{ibm\_torino} quantum processor using 12 qubits for an \( L=6 \) chain, and apply the signal-processing method to improve the measurement results. The correlation analysis (CA) involves measuring \( C^b_{0,k}(t) \) twice, with the raw data shown in Fig.~\ref{fig:Schwinger-result}(\textbf{c}). 
After performing the Fourier transformation, as shown in Fig.~\ref{fig:Schwinger-result}(\textbf{d}), the raw spectra exhibit noise at the low-frequency region near \( E(k) = 0 \), compared with the noiseless Trotter results. When CA is applied alone, the low-frequency noise is significantly suppressed. Moreover, by filtering the two raw spectra with the Hamming window~\cite{4309918} and subsequently applying CA, both the low-frequency noise is reduced and the energy peaks for various momentum values \( k \) align closely with the noiseless Trotter results, as shown in the upper panel of Fig.~\ref{fig:Schwinger-result}(\textbf{e}).

To quantitatively study the hadron spectrum and the effect of the classical signal processing, we fit the energy with the hadron dispersion relation $E(k)^2 = (m_hc^2)^2+(ck)^2$, where $m_h$ is the hadron mass, $c$ is the speed of light on the lattice, as presented by red dashed lines in Fig.~\ref{fig:Schwinger-result}(\textbf{d})~\footnote{The fitting does not exclude the large momentum region or use the lattice dispersion relation, since the continuous dispersion relation is well respected at large $k$ region according to the numerical results}. The lower panel of Fig.~\ref{fig:Schwinger-result}(\textbf{e}) plots the fitted hadron mass with and without the signal processing. The classical processed spectrum using CA and filtering gives the most accurate hadron mass with $m_hc^2|_{\text{CA+filter}}=1.119(36)$, which is in quantitative agreement with the noiseless Trotter one $m_hc^2|_{\text{noiseless}}=1.117(24)$ within a relative error of $0.18\%$. 
Although, these hadron masses are deviated from the  exact diagonalized $m_hc^2|_{\text{ex}}=1.166$ mainly due to the Trotter error.

\begin{figure}
    \centering
    \includegraphics[width=0.47\textwidth]{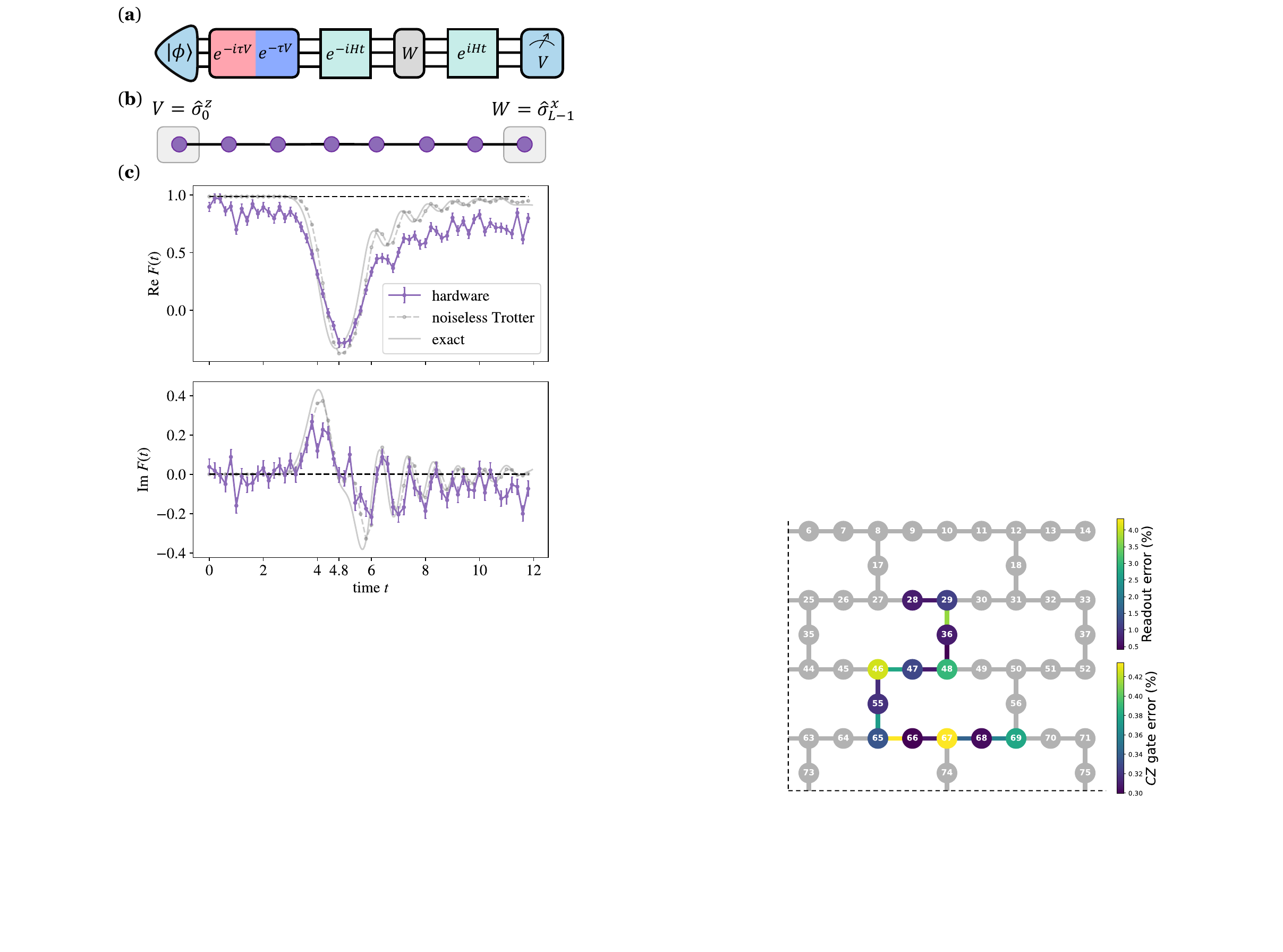}
    \cprotect\caption{Measuring OTOC of the transverse-field Ising model (TIM). (\textbf{a}) The ancilla-free circuits to measure the imaginary part (with the red gate $e^{-i\tau V}$) and the real part (with the blue gate $e^{-\tau V}$) of OTOC. (\textbf{b}) Lattice of TIM and the operator in OTOC. (\textbf{c}) The real and imaginary parts of OTOC $F(t)$ measured on \verb|ibm_torino| versus the time $t$. The grey solid and grey dashed lines denote exact diagonalization and noiseless Trotter results, respectively.}
    \label{fig:otoc-result}
\end{figure}

\vspace{0.2cm}
\noindent\textbf{OTOC ---}
The out-of-time-order correlator (OTOC) is a prob of quantum scrambling defined by the correlation function $F(t)\equiv \bra{\phi} W(t)VW(t)V\ket{\phi}$~\cite{Swingle2018}, where $W(t)=e^{iHt} We^{-iHt}$ is the butterfly operator and $W,V$ are spatially departed Hermitian and unitary operators. We have $F(0)=1$ and the deviation of $F(t)$ from $1$ reflects the scrambling of the quantum system. Here we consider the OTOC of the transverse-field Ising model (TIM) with the Hamiltonian
$
    H_{\text{TIM}} = -\sum_{j=0}^{L-2} \hat{\sigma}^x_{j} \hat{\sigma}^x_{j+1}-\sum_{j=0}^{L-1}\hat{\sigma}^z_j$,
which is integrable by mapping the Hamiltonian to free fermions~\cite{Mbeng_20}.
The real and imaginary parts of $F(t)$ correspond to the 2-time anti-commutator and commutator
\begin{equation}\nonumber
    \begin{aligned}
    \operatorname{Re}F(t) &= \bra{\phi} [W(t)VW(t),V]_+\ket{\phi}/2;\\
    \operatorname{Im}F(t) &= -i \bra{\phi} [W(t)VW(t),V]_-\ket{\phi}/2,
\end{aligned}
\end{equation}
which can be measured using the quantum circuits given by Fig.~\ref{fig:otoc-result}(\textbf{a}). The OTOC of TIM is measured through $V=\hat{\sigma}_0^z$, $W=\hat{\sigma}_{L-1}^x$, and the initial state is $\ket{\phi}=\ket{+}^{\otimes L}$, where $\ket{+}$ is the eigenstate of $\hat{\sigma}_x$ with an eigenvalue $1$. 

In Fig.~\ref{fig:otoc-result}(\textbf{c}), we show the measured real and imaginary parts of \( F(t) \) versus \( t \) on the \( L=8 \) lattice. Here, the Trotter evolution circuits of TIM are compressed using the algebraic compression method~\cite{PhysRevA.105.032420}. The deviation and recurrence of \( \operatorname{Re}F(t) \) from 1 can be observed as time evolves, supporting the integrability of TIM. A significant valley in \( \operatorname{Re}F(t) \) indicates quantum scrambling, with the most scrambled time occurring near \( t=4.8 \). This behavior is consistent with the noiseless Trotter results, denoted by the grey dashed line in Fig.~\ref{fig:otoc-result}(\textbf{c}).

In addition to our method, several quantum simulation experiments have been conducted to measure the out-of-time-order correlator (OTOC) without the use of ancilla qubits~\cite{PhysRevLett.129.160602, Liang2025}. However, these experiments require the initial state \( \ket{\phi} \) to be the eigenstate of the operator \( V \), which allows the time evolution of \( V \) in Fig.~\ref{fig:otoc-result}(\textbf{a}) to be omitted, at the cost of only obtaining the real part of the OTOC. In contrast, our method applies to more general initial states with finite correlation lengths, and measures both the real and imaginary parts of OTOC.

\vspace{0.2cm}
\noindent\textbf{Discussion ---}
In summary, we present a universal approach for measuring $n$-time correlation functions on practical quantum hardware. We apply it to compute the single-particle spectra of the Schwinger model on IBM quantum hardware, accurately measuring the hadron mass with a 0.18\% relative error, in agreement with noiseless Trotter results. We also measure the real and imaginary parts of the OTOC for the transverse-field Ising model. While demonstrated on superconducting qubits, our approach is applicable to general quantum hardware, including both digital and analog quantum simulators.
Our method is comparable to but distinct from existing approaches that eliminate controlled operations~\cite{PhysRevResearch.1.013006, PhysRevLett.132.220601}. Additionally, quantum algorithms using ancilla qubits for measurements, such as variational quantum simulation algorithms~\cite{Yuan2019theoryofvariational, PhysRevLett.125.010501,PhysRevResearch.2.033281}, can benefit from simplifications based on the techniques presented here.



\vspace{0.2cm}
\noindent\textbf{Acknowledgements}
We thank Renbao Liu, Yan Lyu, Yinchenguang Lyu, Ping Wang, Xinhua Peng, Ze Wu, and Li You for insightful discussions and valuable comments. X.L. and X.Y. are supported by the National Natural Science Foundation of China Grant (No.~12361161602), NSAF (Grant No.~U2330201), Beijing Natural Science Foundation Z250004, Beijing Municipal Science and Technology Plan LZ2025-11, Quantum Science and Technology-National Science and Technology Major Project (2023ZD0300200), and the High-performance Computing Platform of Peking University. X.C. is supported by the National Natural Science Foundation of China Grant No.~22303005, and the Startup Fund of the Institute of High Energy Physics, Chinese Academy of Sciences. X.W. is supported by the RIKEN TRIP initiative (RIKEN Quantum) and the University of Tokyo Quantum Initiative. 


\bibliography{main.bib}
\bibliographystyle{IEEEtran} 

\appendix
\section*{End Note}

\textit{Appendix A: Measurement formula for $n$-time correlation functions}~---~ We generalize our discussion on the $2$-time correlation function in the framework section to general $n$-time correlation functions, and show that the $n$-time correlation functions are measured by the ancilla-free circuit in the upper panel of Fig.~\ref{fig:main-circuit}(\textbf{a}). 

The general $n$-time correlation function is $\bra{\phi} O_{n-1}(t_{n-1}) O_{n-2}(t_{n-2}) \cdots O_{1}(t_{1}) O_{0}(t_{0})\ket{\phi}$,
where $O_{j}(t_{j}) = U^{\dagger}(t_j;t_0)O_{j}U(t_j;t_0)$ is the operator evolved in the Heisenberg picture. We assume that $O_{n-1}$ is Hermitian, and $O_{j<n-1}$ are unitary and Hermitian Pauli strings. Non-unitary and non-Hermitian operators can be expanded as linear combinations of Pauli strings, and thus can also be addressed using our method.

We denote the nested commutators and anti-commutators:
\begin{equation}
    \begin{aligned}
    c_{\bos{b}}\equiv \bra{\phi}[[\cdots[O_{n-1}&(t_{n-1}),O_{n-2}(t_{n-2})]_{b_{n-2}},\\
    &\ldots ,O_1(t_1)]_{b_{1}},O_0(t_0)]_{b_{0}} \ket{\phi},
    \label{eq:nested-commutators-anticommutators}
\end{aligned}
\end{equation}
where $\bos{b}\equiv (b_0,b_1,\ldots,b_{n-2}),~b_j=+,-$ is an $(n-1)$-tuple. Then, the $n$-time correlation function can be expressed as a linear combination of $c_{\bos{b}}$. For example, $\bra{\phi}O_1(t_1)O_0(t_0)\ket{\phi} =\frac{1}{2}(c_++c_-)$. Using this relation iteratively, all the $n$-time correlation functions can be written as a linear combination of $2^{n-1}$ nested commutators and anti-commutators, which is efficient as long as $n=\OO(1)$.

Next, the nested commutators and anti-commutators $c_{\bos{b}}$ can be derived from the partial derivatives of the circuit expectation $\langle O_{n-1}\rangle_{\bos{b}}(\bos{\tau})$ in the upper panel of Fig.~\ref{fig:main-circuit}(\textbf{a}). For each $c_{\bos{b}}$, we define the unitary real-time and imaginary-time evolution of $O_j$ in Fig.~\ref{fig:main-circuit}(\textbf{a}) as 
\begin{align}
    U_{O_j}^{(b_j)}(\tau_j)\ket{\psi_0}\equiv\left\{\begin{array}{ll}
 e^{-i\tau_j O_j}\ket{\psi_0} &\text{for } b_j=-;\\
 \frac{e^{-\tau_j O_j}}{\sqrt{\bra{\psi_0}e^{-2\tau_j O_j}\ket{\psi_0}}}\ket{\psi_0} &\text{for } b_j=+,
  \end{array}\right.
  \label{eq:stimulate-unitaries}
\end{align}
with an input state $\ket{\psi_0}$. Using the composition property $U(t_j; t_{j-1})=U(t_j;t_0)U^{\dagger}(t_{j-1};t_0)$, it can be shown that each $O_j$ is evolved to the desired time $t_j$, such that $U_{O_j}^{(b_j)}$ in Fig.~\ref{fig:main-circuit}(\textbf{a}) is evolved to $U_{O_j(t_j)}^{(b_j)}$. To measure $c_{\bos{b}}$ using the expectation $\langle O_{n-1}\rangle_{\bos{b}}(\bos{\tau})$, we first ignore the normalization factor in $U_{O_j(t_j)}^{(+)}$ for simplicity, and notice that the commutator and anti-commutator of an arbitrary operator $A$ equal the partial derivatives
\begin{equation}
    \begin{aligned}
    [A, O_j(t_j)]_- &= i\partial_\tau (e^{i\tau O_j(t_j)}A e^{-i\tau O_j(t_j)})|_{\tau=0};\\
    [A, O_j(t_j)]_+ &= -\partial_\tau (e^{-\tau O_j(t_j)}A e^{-\tau O_j(t_j)})|_{\tau=0}.
\end{aligned}
\end{equation}
Thus, $c_{\bos{b}}$ with multiple commutators and anti-commutators is derived by the following high-order derivative evaluated at $\bos{\tau}=\bos{0}$
\begin{align}
    c_{\bos{b}} = (-1)^{n_{+,\bos{b}}} (i)^{n_{-,\bos{b}}}\partial_{\tau_{0}}\cdots \partial_{\tau_{n-2}} \langle  O_{n-1}\rangle_{\bos{b}}|_{\bos{\tau=\bos{0}}},
    \label{eq:correlation-function-by-derivatives}
\end{align}
where $n_{+,\bos{b}}$ and $n_{-,\bos{b}}$ are the numbers of plus and minus signs in $\bos{b}$. This gives the key formula for measuring $n$-time nested commutators and anti-commutators using the ancilla-free circuit.

For the normalization factor in $U_{O_j}^{(+)}(\tau_j)$, we use the 2-time correlation function as an example, with a full discussion in Supplemental Material~\cite{supp}. The true derivative of the circuit expectation $\langle O_1\rangle_+(\tau)$ in Fig.~\ref{fig:main-circuit}(\textbf{d}) gives:
\begin{align}
    \partial_{\tau}\langle O_1\rangle_+  |_{\tau=0}= -c_+ +2\bra{\phi}O_1(t_1)\ket{\phi}\bra{\phi}O_0\ket{\phi},
    \label{eq:differential-correction-formula}
\end{align}
with an additional term $\bra{\phi} O_1(t_1) \ket{\phi} \bra{\phi} O_0 \ket{\phi}$. To obtain the anti-commutator $c_+$, we subtract this term from the partial derivative using additional measurement circuits that are no more complicated than the main circuit in Fig.~\ref{fig:main-circuit}(\textbf{d}).

\textit{Appendix B: Parameter-shift rule}~---~ This section presents the main formula for evaluating the $n$-time nested commutators and anti-commutators $c_{\bos{b}}$ using the parameter-shift rule, considering the effects of the QITE normalization factor.

The parameter-shift rule evaluates each partial derivative with respect to the real or imaginary evolution time at two time values. Thus, $c_{\bos{b}}$ with $(n-1)$ partial derivatives in Eq~\eqref{eq:correlation-function-by-derivatives} should be evaluated at $2^{n-1}$ time points indexed by a binary tuple $\bos{\xi}$:
\begin{widetext}
\begin{equation}
    \begin{aligned}
    c_{\bos{b}} = (-1)^{n_{+,\bos{b}}}  (i)^{n_{-,\bos{b}}}\sum_{\bos{\xi}\in\{0,1\}^{n-1}} &(-1)^{|\bos{\xi}|} \text{Corr}((-1)^{\xi_0}\tau_{b_0},\ldots,(-1)^{\xi_{\ii}}\tau_{b_{\ii}})
    \times \langle O_{n-1}\rangle_{\bos{b}} (\bos{\tau}=((-1)^{\xi_0}\tau_{b_0},\ldots, (-1)^{\xi_{n-2}}\tau_{b_{n-2}})),
    \label{eq:corrected-c_b-by-parameter-shift-rule-note}
\end{aligned}
\end{equation}
\end{widetext}
where $\bos{\xi}$ takes all the $(n-1)$-site binary numbers with $|\bos{\xi}|=\sum_j \xi_j$ as its Hamming weight. Since $b_j=\pm$,  $\tau_{\pm}$ in the tuple $\bos{\tau}$ are the imaginary and real evolution time taking values $\tau_+=\frac{1}{2}\log(1+\sqrt{2})$ and $\tau_-=\pi/4$ for the normalization: $\sinh 2\tau_+=\sin2\tau_-=1$.  $\text{Corr}((-1)^{\xi_0}\tau_{b_0},\ldots,(-1)^{\xi_{\ii}}\tau_{b_{\ii}})$ is a correction factor comes from the normalization factor in $U_{O_j}^{(+)}$, as mentioned in Appendix A. Here, $\ii\equiv\max\{~j~|~b_j=+\}$ is the index of $U_{O_{j=\ii}}^{(+)}$ closest to the observable $O_{n-1}$ in Fig~\ref{fig:main-circuit}(\textbf{a}). If $c_{\bos{b}}$ is a nested commutator and no $U_{O_j}^{(+)}$ is involved, then $\text{Corr}=1$. Otherwise, the correction factor is evaluated according to its definition 
\begin{equation}
    \begin{aligned}
    \text{Corr}(\tau_0,\ldots,\tau_{\ii})\equiv \bra{\phi} P^{(b_0)\dagger}_{O_0(t_0)}\cdots P^{(b_{\ii-1})\dagger}_{O_{\ii-1}(t_{\ii-1})}e^{-2\tau_{\ii}O_{\ii}}\\
    P^{(b_{\ii-1})}_{O_{\ii-1}(t_{\ii-1})}\cdots P^{(b_0)}_{O_0(t_0)}\ket{\phi},
    \label{eq:Corr-formula-app}
\end{aligned}
\end{equation}
where $P_{O_j(t_j)}^{(b_j)}$ are real and imaginary-time propagators that can be expanded using Euler's identity
\begin{equation}
    \begin{aligned}
    P_{O_j(t_j)}^{(-)}&\equiv e^{-i\tau_jO_j(t_j)}=\cos\tau_j-i\sin\tau_j~O_j(t_j);\\
    P_{O_j(t_j)}^{(+)}&\equiv e^{-\tau_jO_j(t_j)}=\cosh\tau_j-\sinh\tau_j~O_j(t_j).
\end{aligned}
\end{equation}
\noindent The correction factor contains at most $(n-1)$-time correlation function. Thus, one can use Eq.~\eqref{eq:corrected-c_b-by-parameter-shift-rule-note} iteratively to evaluate the correction factor without ancilla qubits. The derivation of these formulae is detailed in Supplemental Material~\cite{supp}.

\begin{figure}
    \centering
\includegraphics[width=0.35\textwidth]{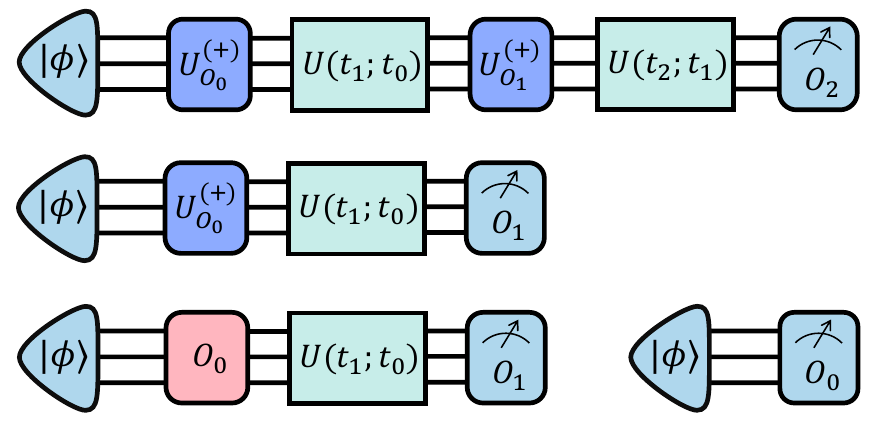}
    \caption{Quantum circuits to measure the 3-time nested anti-commutator $\bra{\phi}[[O_2(t_2),O_1(t_1)]_+,O_0(t_0)]_+\ket{\phi}$ and its correction factor.}
    \label{fig:three-time-circuits}
\end{figure}

As an example, the 3-time nested anti-commutator $\bra{\phi}[[O_2(t_2),O_1(t_1)]_+,O_0(t_0)]_+\ket{\phi}$ has the correction factor
\begin{equation}
    \begin{aligned}
    \text{Corr}(\tau_0,\tau_1) =&\cosh 2\tau_1[\cosh 2\tau_0 -\sinh2\tau_0\bra{\phi} O_0(t_0)\ket{\phi}]\\
    &- \sinh 2\tau_1 \{\cosh^2\tau_0\bra{\phi} O_1(t_1)\ket{\phi}\\
    &-\cosh\tau_0\sinh\tau_0 \bra{\phi} [O_1(t_1),O_0(t_0)]_+\ket{\phi}\\
    &+\sinh^2\tau_0\bra{\phi} O_0(t_0)O_1(t_1)O_0(t_0)\ket{\phi} \}.
\end{aligned}
\end{equation}
This factor contains the 2-time anti-commutator $\bra{\phi} [O_1(t_1), O_0(t_0)]_+\ket{\phi}$ that can be evaluated as introduced in the main text. The last term $\bra{\phi} O_0(t_0)O_1(t_1)O_0(t_0)\ket{\phi}$ is symmetric in time $t_0$ and thus can be regarded as a $2$-time correlation function.  Figure~\ref{fig:three-time-circuits} illustrates all the quantum circuits to measure the 3-time nested anti-commutator and its correction factor. The lower three circuits are utilized to evaluate the correction factor that are no more complicated than the main circuit in the first row of Fig.~\ref{fig:three-time-circuits}.

\onecolumngrid
\newpage

\setcounter{page}{1}
\renewcommand{\thefigure}{S\arabic{figure}}
\renewcommand{\theequation}{S\arabic{equation}}


\titleformat{\section}{\normalfont\large\bfseries}{Supplementary Note \thesection:}{1em}{}

\section{Parameter-shift rule for $n$-time correlation functions}\label{app:parameter-shift-rule}

In this note, we first present how the nested commutator and anti-commutator $c_{\bos{b}}$ is derived from the partial derivatives in Eq.~(6) of the End Note. To uniformly describe the real-time and imaginary-time evolution of operators, for each $(n-1)$-tuple $\bos{b}=(b_0,\ldots, b_{n-2})$, we introduce $\bos{\zeta}=(\zeta_0,\ldots, \zeta_{n-2})$ with $\zeta_j=1$ if $b_j=+$, and $\zeta_j=i$ if $b_j=-$. With these notations, using the composition property $U(t_j; t_{j-1})=U(t_j;t_0)U^{\dagger}(t_{j-1};t_0)$, the expectation $\langle O_{n-1}\rangle_{\bos{b}}$ reads
\begin{align}
    \langle O_{n-1}\rangle_{\bos{b}}(\bos{\tau}) = \bra{\phi}e^{-\bar{\zeta}_0\tau_0 O_0(t_0)}\ldots e^{-\bar{\zeta}_{n-2}\tau_{n-2} O_{n-2}(t_{n-2})} O_{n-1}(t_{n-1}) e^{-\zeta_{n-2}\tau_{n-2} O_{n-2}(t_{n-2})}\ldots e^{-\zeta_{0}\tau_{0} O_{0}(t_{0})}\ket{\phi},
\end{align}
where $\overline{(\cdot)}$ denotes complex conjugation. Its derivative with respect to $\tau_{n-2}$ gives
\begin{align}
    \partial_{\tau_{n-2}}\langle O_{n-1}\rangle_{\bos{b}}(\bos{\tau})|_{\tau_{n-2}=0} =-\zeta_{n-2}\bra{\phi}e^{-\bar{\zeta}_0\tau_0 O_0(t_0)}\ldots [O_{n-1}(t_{n-1}),O_{n-2}(t_{n-2})]_{b_{n-2}}
    \ldots e^{-\zeta_{0}\tau_{0} O_{0}(t_{0})}\ket{\phi}.
\end{align}
The commutator or anti-commutator appears by the derivative for $b_{n-2}=-$ or $+$. Thus, taking all the partial derivatives with respect to $\tau_{n-2},\ldots,\tau_0 $ gives 
\begin{align}
    \partial_{\tau_{0}}\ldots \partial_{\tau_{n-2}}\langle O_{n-1}\rangle_{\bos{b}}(\bos{\tau})|_{\bos{\tau}=\bos{0}} =(-1)^{n-1}\left(\prod_{j=0}^{n-2}\zeta_{j}\right)  c_{\bos{b}},
\end{align}
which is equivalent to the key formula Eq.~(6) in the End Note. Numerically, the partial derivatives can be evaluated using a naive finite-difference approach. However, the statistical and truncation errors in the final measurement results are amplified by the denominator in the finite-difference approach.

Instead, if $O_{j}$ is an Hermitian and unitary Pauli string, the corresponding partial derivative can be evaluated accurately using the parameter-shift rule, which states that in a quantum circuit with a rotation gate $e^{-i\tau O}$, if $O$ is a Pauli string, the derivatives of the circuit expectation with respect to $\tau$ is proportional to the difference of two circuits with shifted parameters. For example, the measurement circuit in Fig.~1(\textbf{c},\textbf{d}) of the main text has the Pauli string $O_0$ respecting $O_0O_0=I$ (and also $O_0(t)O_0(t)=I, \forall t\in \mathbb{R}$). Since the exponential can be expressed using Euler's identity $e^{-i\tau_0O_0(t_0)} = \cos\tau_0 -i\sin\tau_0~O_{0}(t_0)$ and $e^{-\tau_0O_0(t_0)} = \cosh\tau_0 -\sinh\tau_0~O_{0}(t_0)$, the 2-time commutator and anti-commutator can be evaluated through
\begin{equation}
    \begin{aligned}
    \bra{\phi}[O_1(t_1),O_0(t_0)]_-\ket{\phi}&=i\frac{1}{\sin 2\tau} [\langle O_1\rangle_-(\tau)-\langle O_1\rangle_-(-\tau)];\\
    \bra{\phi}[O_1(t_1),O_0(t_0)]_+\ket{\phi} &=-\frac{1}{\sinh 2\tau} [\langle O_1\rangle_+(\tau)-\langle O_1\rangle_+(-\tau)].
    \label{eq:partile-derivative-by-parameter-shift-rule}
\end{aligned}
\end{equation}
where $\tau$ is an arbitrary real number. Combining these two equations gives the parameter-shift rule to evaluate arbitrary $c_{\bos{b}}$:
\begin{equation}
    \begin{aligned}
    c_{\bos{b}} =\left(\frac{-1}{\sinh 2\tau_-}\right)^{n_{+,\bos{b}}}\left(\frac{i}{\sin 2\tau_+}\right)^{n_{-,\bos{b}}}\sum_{\bos{\xi}\in\{0,1\}^{n-1}} (-1)^{|\bos{\xi}|} \langle O_{n-1}\rangle_{\bos{b}}(\bos{\tau}=((-1)^{\xi_0}\tau_{b_{0}}, \ldots, (-1)^{\xi_{n-2}}\tau_{b_{n-2}})),
    \label{eq:c_b-by-parameter-shift-rule}
\end{aligned}
\end{equation}
where $\tau_-$ and $\tau_+$ are arbitrary real number as the real and imaginary evolution time, respectively. $n_{+,\bos{b}}$ and $n_{-,\bos{b}}$ are the numbers of plus and minus signs in the $(n-1)$-tuple $\bos{b}$. $\bos{\xi}\in\{0,1\}^{n-1}$ takes all the $(n-1)$-site binary numbers, $|\bos{\xi}|=\sum_j \xi_j$ is $\bos{\xi}$'s Hamming weight.

We have three comments on Eq.~\eqref{eq:c_b-by-parameter-shift-rule}. First, using the parameter-shift rule, for $n$-time nested commutators and anti-commutators, we need $2^{n-1}$ measurements to obtain $c_{\bos{b}}$. However, a small $n\sim\OO(1)$ is enough for theoretical prediction in many relevant physical applications. Second, for the practical use of Eq.~\eqref{eq:c_b-by-parameter-shift-rule}, $\tau_-,\tau_+\sim\OO(1)$ is preferable such that the errors in the $O_{n-1}$ expectation are not amplified by the small $\sinh 2\tau_+$ and $\sin 2\tau_-$ in Eq.~\eqref{eq:c_b-by-parameter-shift-rule}. Additionally, $\tau_+$ should not be too large, since the gate counts of the quantum imaginary-time evolution (QITE) grows with $\tau$, as discussed in Supplementary Note~\ref{app:high-order-qite}. For example, we can choose $\tau_-=\pi/4$ and $\tau_+=\frac{1}{2}\log(1+\sqrt{2})$, such that $\sinh 2\tau_+=\sin 2\tau_-=1$. Third, due to the normalization factor of QITE, the second line of Eq.~\eqref{eq:partile-derivative-by-parameter-shift-rule} has an additional correction term. The corrected formula measuring 2-time anti-commutator reads
\begin{equation}
    \begin{aligned}
    \bra{\phi}[O_1(t_1),O_0(t_0)]_+\ket{\phi} = -\frac{1}{\sinh 2\tau_+}[&(\cosh 2\tau_+ -\sinh 2\tau_+ \bra{\phi}O_0\ket{\phi})~\langle O_1\rangle_+(\tau_+)\\
    &-(\cosh 2\tau_+ +\sinh 2\tau_+ \bra{\phi}O_0\ket{\phi})~\langle O_1\rangle_+(-\tau_+)].
    \label{eq:2-time-anticommutator-corrected}
\end{aligned}
\end{equation}
Taking the limit $\tau_+\to 0$ gives rise to Eq.~(7) in the End Note. Compared with Eq.~\eqref{eq:partile-derivative-by-parameter-shift-rule}, QITE normalization factor requires additional measuring the expectation $\bra{\phi}O_0\ket{\phi}$. A more general corrected version of  Eq.~\eqref{eq:c_b-by-parameter-shift-rule} will be discussed in the remainder of this note.
\\\\
\textbf{Correction of Eq.~\eqref{eq:c_b-by-parameter-shift-rule} by QITE normalization factor}~---~Recall that QITE of an observable $O_j$ with an initial state $\ket{\psi}$ reads
\begin{align}
    U_{O_j}^{(+)}(\tau_j)\ket{\psi}=\frac{e^{-\tau_j O_j}}{\sqrt{\bra{\psi}e^{-2\tau_j O_j}\ket{\psi}}}\ket{\psi}.
  \label{eq:stimulate-unitaries-app}
\end{align}
To address the effect of the normalization factor $\sqrt{\bra{\psi}e^{-2\tau_j O_j}\ket{\psi}}$, we firstly investigate the QITE closest to the observable $O_{n-1}$. For a particular $\bos{b}=(b_0,\ldots,b_{n-2})$, we assume that the QITE closest to $O_{n-1}$ is indexed by $\ii\equiv\max\{~j~|~b_j=+\}$, such that 
\begin{align}
    \bos{b}=(b_0,\ldots,b_{\ii-1},b_\ii=+, b_{\ii+1}=-,\ldots,b_{n-2}=-).
\end{align}
The real-time evolutions characterized by $b_{\ii+1}=-,\ldots,b_{n-2}=-$ do not include the normalization factor. We can firstly derive the nested commutators by the parameter-shift rule of these real-time evolutions
\begin{equation}
    \begin{aligned}
    I(\tau_0,\ldots,\tau_{\ii})\equiv&\frac{\bra{\phi} U^{(b_0)\dagger}_{O_0(t_0)}\ldots U^{(b_{\ii-1})\dagger}_{O_{\ii-1}(t_{\ii-1})}e^{-\tau_\ii O_\ii(t_{\ii})} \mathbb{C}(n-1;\ii+1)e^{-\tau_\ii O_\ii(t_{\ii}) }U^{(b_{\ii-1})}_{O_{\ii-1}(t_{\ii-1})}\ldots U^{(b_0)}_{O_0(t_0)}\ket{\phi}}{\bra{\phi} U^{(b_0)\dagger}_{O_0(t_0)}\ldots U^{(b_{\ii-1})\dagger}_{O_{\ii-1}(t_{\ii-1})}e^{-2\tau_\ii O_\ii(t_{\ii})}U^{(b_{\ii-1})}_{O_{\ii-1}(t_{\ii-1})}\ldots U^{(b_0)}_{O_0(t_0)}\ket{\phi}}\\
    =& \left(\frac{i}{\sin 2\tau}\right)^{n-2-\ii}\sum_{\bos{\xi'}\in\{0,1\}^{n-2-\ii}}(-1)^{|\bos{\xi'}|}\langle O_{n-1}\rangle (\bos{\tau}=(\tau_0,\ldots, \tau_{\ii}, \tau_{\ii+1}=(-1)^{\xi'_0}\tau,\ldots, \tau_{n-2}=(-1)^{\xi'_{n-3-\ii}}\tau)),
    \label{eq:I-definition}
\end{aligned}
\end{equation}
where $\tau$ is an arbitrary real number, and we define
\begin{align}
    \mathbb{C}(n-1;\ii+1)\equiv [[\ldots[O_{n-1}(t_{n-1}),O_{n-2}(t_{n-2})]_-,\ldots ]_-,O_{\ii+1}(t_{\ii+1})]_-
\end{align}
in case of $\ii<n-2$, and 
\begin{align}
    \mathbb{C}(n-1;n-1)\equiv O_{n-1}(t_{n-1})
\end{align}
in case of $\ii=n-2$.

Next, we expand the imaginary-time evolution in the first line of Eq.~\eqref{eq:I-definition}. Notice that in Eq.~\eqref{eq:I-definition}, all the unitary $U_{O_j(t_j)}^{(b_j)}(\tau_j),~0 \leq j\leq \ii-1 $ and its conjugation appear once in both the numerator and denominator. Thus, we can substitute the unitary imaginary-time evolutions with the imaginary-time propagators
\begin{align}
    U_{O_j(t_j)}^{(+)}(\tau_j)\rightarrow P^{(+)}_{O_j(t_j)}(\tau_j) \equiv e^{-\tau_j O_j(t_j)}, j<\ii
\end{align}
in both the numerator and denominator of $I(\tau_0,\ldots,\tau_{\ii})$ without changing its value. For convenience, we also define the real-time propagator $P^{(-)}_{O_j(t_j)}(\tau_j) \equiv e^{-i\tau_j O_j(t_j)}=U^{(-)}_{O_j(t_j)}(\tau_j)$. This substitution gives
\begin{align}
    I(\tau_0,\ldots,\tau_{\ii})= \frac{\bra{\phi} P^{(b_0)\dagger}_{O_0(t_0)}\ldots P^{(b_{\ii-1})\dagger}_{O_{\ii-1}(t_{\ii-1})}e^{-\tau_\ii O_\ii(t_{\ii})} \mathbb{C}(n-1;\ii+1)e^{-\tau_\ii O_\ii(t_{\ii}) }P^{(b_{\ii-1})}_{O_{\ii-1}(t_{\ii-1})}\ldots P^{(b_0)}_{O_0(t_0)}\ket{\phi}}{\text{Corr}(\tau_0,\ldots,\tau_{\ii})},
    \label{eq:I-propator}
\end{align}
where
\begin{equation}
    \begin{aligned}
    \text{Corr}(\tau_0,\ldots,\tau_{\ii})\equiv \bra{\phi} P^{(b_0)\dagger}_{O_0(t_0)}\ldots P^{(b_{\ii-1})\dagger}_{O_{\ii-1}(t_{\ii-1})}e^{-2\tau_{\ii}O_{\ii}}P^{(b_{\ii-1})}_{O_{\ii-1}(t_{\ii-1})}\ldots P^{(b_0)}_{O_0(t_0)}\ket{\phi}.
    \label{eq:Corr-formula}
\end{aligned}
\end{equation}
Since the numerator of Eq.~\eqref{eq:I-propator} has no normalization factor, we can derive $c_{\bos{b}}$ from the parameter-shift rule of the numerator, which equals to $\text{Corr}(\tau_0,\ldots,\tau_{\ii})\times I(\tau_0,\ldots,\tau_{\ii})$. Finally, we have
\begin{equation}
    \begin{aligned}
    c_{\bos{b}} = \left(\frac{-1}{\sinh 2\tau_{+}}\right)^{n_{+,\bos{b}}}\left(\frac{i}{\sin 2\tau_-}\right)^{n_{-,\bos{b}}}\sum_{\bos{\xi}\in\{0,1\}^{n-1}} (-1)^{|\bos{\xi}|} \text{Corr}((-1)^{\xi_0}\tau_{b_0},\ldots,(-1)^{\xi_{\ii}}\tau_{b_{\ii}})\\
    \times \langle O_{n-1}\rangle_{\bos{b}} (\bos{\tau}=((-1)^{\xi_0}\tau_{b_0},\ldots, (-1)^{\xi_{n-2}}\tau_{b_{n-2}})).
    \label{eq:corrected-c_b-by-parameter-shift-rule}
\end{aligned}
\end{equation}
Taking $\tau_-=\pi/4$ and $\tau_+=\frac{1}{2}\log(1+\sqrt{2})$ gives Eq.~(8) in the End Note. Compared with Eq.~\eqref{eq:c_b-by-parameter-shift-rule}, the correction factor $\text{Corr}((-1)^{\xi_0}\tau_{b_0},\ldots,(-1)^{\xi_{\ii}}\tau_{b_{\ii}})$ is multiplied to the expectation. The correction factor in Eq.~\eqref{eq:Corr-formula} can be expanded using Euler's identity and measured using similar but shorter circuits than the main circuit of measuring $\langle O_{n-1}\rangle_{\bos{b}}$. For example, the 2-time anti-commutator has the correction factor
\begin{align}
    \text{Corr}(\tau) =\cosh 2\tau- \sinh 2\tau \bra{\phi} O_{0}(t_0)\ket{\phi}.
\end{align}
Taking this correction factor to Eq.~\eqref{eq:corrected-c_b-by-parameter-shift-rule} gives rise to the corrected measurement formula Eq.~\eqref{eq:2-time-anticommutator-corrected}.

\section{Bang-bang circuits for digital-analog platforms}\label{app:DAQC-realization}

\begin{figure}
    \centering
    \includegraphics[width=0.5\textwidth]{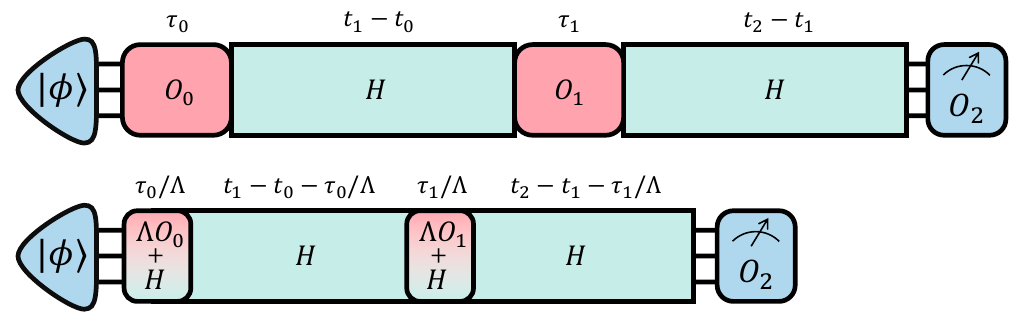}
    \caption{Measurement circuits for $3$-time nested commutator (upper panel) and its bang-bang version suitable for digital-analog platforms (lower panel). $H$ is the native Hamiltonian of the analog platform. $\Lambda$ is a rescaling factor. The evolution time of each block is labeled above each block. Compared with the original circuit, the bang-bang circuit has the system evolution on during the whole time.}
    \label{fig:banged-circuit}
\end{figure}

Our measurement circuit for the $n$-time correlation function in Fig.~1(\textbf{a}) can be realized straightforwardly on digital-analog platforms with a robust unitary evolution $U(t_i; t_j)$. However, it requires turning on and off the unitary evolution infinitely fast which is practically unfeasible. Instead, we introduce the bang-bang measurement circuit without turning off the evolution, as illustrated in Fig.~\ref{fig:banged-circuit}. Here, we use $H$ to denote the native Hamiltonian of the analog platform, and a rescaling factor $\Lambda\gg1$ to reduce the error, as will be discussed later. Compared with the original circuit in the upper panel of Fig.~\ref{fig:banged-circuit}, the bang-bang circuit in the lower panel has the native evolution on during the whole time. The operators are applied using bang-bang pulses~\footnote{Here ``bang-bang'' means turning on and off in a short interval.}, such that the whole evolution is described by a time-dependent Hamiltonian 
\begin{align}
    H'(t) = H+\sum_{j=0}^{n-2} \Lambda O_j \left[\theta(t-t_j)- \theta(t-(t_j+\tau_j/\Lambda))\right].
\end{align}
These rectangular pulses are similar to the dynamic decoupling pulses frequently used to suppress decoherence in quantum control~\cite{PhysRevA.58.2733}. In practice, the rectangular window can be substituted by other window functions that are more feasibly realized. In Fig.~\ref{fig:banged-circuit}, we use the 3-time nested commutator as an example. The circuit can be generalized directly to the case of $n$-time nested commutator and anti-commutators. 

The bang-bang circuit introduces additional error to the measurement results that can be controlled by $\Lambda$. The error from one pulse with $\Lambda\gg 1$ and $\tau_j\sim 1$ can be estimated using Wilcox formula~\cite{10.1063/1.1705306}
\begin{equation}
    \begin{aligned}
    \epsilon_j\equiv \|e^{-iH\tau_j/\Lambda}e^{-iO_j\tau_j} - e^{-i(H+\Lambda O_j)\tau_j/\Lambda}\|&= \|\frac{-i\tau_j}{\Lambda}[H-\int_{0}^{1}\ud s~e^{-is\tau_j O_j} H e^{is\tau_j O_j}]e^{-iO_j\tau_j}+\OO(1/\Lambda^2)\|\\
    &\leq \frac{|\tau_j|}{\Lambda} \|H\|(1+\frac{|\sin 2\tau_j|}{|2\tau_j|}+\frac{|1-\cos 2\tau_j|}{|2\tau_j|})  +\OO(1/\Lambda^2)\\
    &\leq \frac{2.31|\tau_j|}{\Lambda} \|H\|+\OO(1/\Lambda^2),
\end{aligned}
\end{equation}
where we explicitly performed the integration in the second line and used the fact that $O_j$ is a Pauli string with the spectral norm $\|O_j\|=1$. We see that the error from one bang-bang pulse is linearly suppressed by $\Lambda$, and so as the total error from the bang-bang measurement circuits $E=\sum_j \epsilon_j =\OO(n/\Lambda)$. Thus, the bang-bang error in the final measurement results can be suppressed by reducing the bang-bang pulse duration and increasing its intensity correspondingly.

\section{Quantum imaginary-time evolution for Pauli strings with finite evolution time}\label{app:high-order-qite}

In this note, we introduce the realization of QITE with a finite evolution time $\tau$ required by the parameter-shift rule and estimate the required gate counts. Assume that an operator $O=\widetilde{\sigma}$ is a Pauli string acting on at most $k$ neighboring qubits. Its imaginary-time evolution for the initial state $\ket{\psi_0}$ can be realized by firstly Trotter decomposed as
\begin{align}
    e^{-\tau\widetilde{\sigma}}\ket{\psi_0}=(e^{-\Delta \tau \widetilde{\sigma}})^r\ket{\psi_0};~ r=\frac{\tau}{\Delta \tau}.
\end{align}
Then, QITE performs each Trotter step by unitary gates $U_m,m\in[1,r]$. We define the evolved state after $m\in [0, r]$ steps as 
\begin{align}
    \ket{\psi_m}\equiv\frac{e^{-m\Delta \tau\widetilde{\sigma}}}{\sqrt{\bra{\psi_0}e^{-2m\Delta\tau\widetilde{\sigma}}\ket{\psi_0}}}\ket{\psi_0}=U_m\ldots U_2U_1\ket{\psi_0}.
\end{align}

Each unitary $U_m$ is realized as follows. If $\ket{\psi_0}$ has a finite correlation length, then all $U_m$ are local unitary transformations acting on the neighboring qubits of $\widetilde{\sigma}$ (including the qubits that $\widetilde{\sigma}$ acting on). We denote the set of these local qubits as $R$, the number of qubits in $R$ will be rigorously evaluated in the following subsection. Any unitaries $U_m$ on $R$ can be expanded by 
\begin{align}
    U_m = e^{-i \Delta\tau \sum_l  a_l\sigma_l},
    \label{eq:QITE-unitary}
\end{align}
where $\sigma_l$ in general includes all Pauli strings on $R$, and $a_l$ can be obtained by solving the linear system of equations~\cite{Motta_20}
\begin{align}
    \sum_{l'} M_{l,l'} a_{l'} = V_{l},
    \label{eq:DetQITE-linear-equation}
\end{align}
where 
\begin{equation}
\begin{aligned}
    M_{l,l'}\equiv \Re(\bra{\psi_{m-1}}\sigma_{l}\sigma_{l'}\ket{\psi_{m-1}});~V_{l}\equiv \Im(\bra{\psi_{m-1}}\sigma_l \widetilde{\sigma}\ket{\psi_{m-1}}).
    \label{eq:M-V-calculation-measure}
\end{aligned}
\end{equation}
Since the product of two Pauli strings is still a Hermitian Pauli string up to a prefactor, both the $M$ matrix and $V$ vector can be calculated by evaluating the expectations on quantum computers without ancilla qubits. 

Finally, the summation in Eq.~\eqref{eq:QITE-unitary} can be converted into a product of Pauli exponentials $\prod_l e^{-i\Delta \tau a_l\sigma_l}$, and each Pauli exponential can be implemented using standard unitary gates. In the following two subsections, we estimate the total gate counts of QITE and provide two examples of QITE used in our hardware demonstrations.

\subsection{Gate counts bounds of the quantum imaginary-time evolution}\label{app:Run time bounds of the quantum imaginary-time evolution}
In this subsection, we estimate the gate counts of QITE of the operator $O=\widetilde{\sigma}$. Firstly, we estimate the number of qubits in $R$ that the unitaries $U_m$ acts on, such that the whole QITE can be realized as 
\begin{align}
   \left\| \frac{e^{-\tau \widetilde{\sigma}}}{\sqrt{\bra{\psi_0} e^{-2\tau \widetilde{\sigma}}\ket{\psi_0}}}\ket{\psi_0}- U_r\ldots U_1 \ket{\psi_0}\right\| \leq \epsilon , 
    \label{eq:qite-single-step}
\end{align}
Here, we assume that $\widetilde{\sigma}$ has weight $k$ acting on a $d$-dimensional lattice, which is frequently met in physical applications, and the initial state $\ket{\psi_0}$ has an upper-bounded correlation length $\xi$, i.e., for every pair of local operator $A, B$ on the $d$-dimensional lattice separated by a Euclidean distance $d(A, B)$, their connected correlation decays exponentially as 
\begin{align}
    |\bra{\psi_0} AB\ket{\psi_0}-\bra{\psi_0} A\ket{\psi_0}\bra{\psi_0} B\ket{\psi_0}|\leq \|A\|\|B\| e^{-d(A,B)/\xi}.
\end{align}
Then, similar to the proof of Theorem 1 in Ref.~\cite{Motta_20}, using Uhlmanns' theorem, one can show that there exists a unitary $\hat{U}$ acting on a local region $R$ of the $d$-dimensional lattice with diameter $v$, such that
\begin{align}
    \left\| \frac{e^{-\tau \widetilde{\sigma}}}{\sqrt{\bra{\psi_0} e^{-2\tau \widetilde{\sigma}}\ket{\psi_0}}}\ket{\psi_0}- \hat{U}\ket{\psi_0}\right\| \leq 2e^{2\tau-v/(2\xi)}.
\end{align}
Thus, choosing the diameter $v = 2\xi(2\tau+ \ln (2/\epsilon))$, we can control the above error under $\epsilon$. The number of qubits in this $d$-dimensional local region $R$ is at most 
\begin{align}
    N_q= k [2\xi(2\tau+ \ln (2/\epsilon))]^d,
\end{align}
and the number of Pauli strings in the expansion of $U_m$ in Eq.~\eqref{eq:QITE-unitary} scales as $4^{\OO(N_q)}$. This is also the cost scaling of solving the linear system of equations for $a_l$. Thus, the gate counts of QITE is bounded by
\begin{align}
    N_G =  4^{\OO(k [2\xi(2\tau+ \ln (2/\epsilon))]^d)}.
    \label{eq:running-time-scaling}
\end{align}
For a constant precision $\epsilon$ and an $\OO(1)$ evolution time $\tau$ required by the parameter-shift rule, this gate counts has no dependence on the system size. Thus, one QITE requires gate counts of $4^{\OO(k\xi^d)}$.

Compared with the Hadamard test method, the ancilla-free measurement protocol saves the gate counts by considering the finite qubit connectivity of current devices. Assume that the device connectivity is a $d$-dimensional lattice with $N$ qubits. For the operator $O$ acting on $k$ qubits with $k\ll N$, the Hadamard test swaps these $k$ qubits to the control qubit to perform a local controlled operation. The swap process requires additional gate counts of $\OO(kN^{1/d})$. Combined with Eq.~\eqref{eq:running-time-scaling}, measuring the anti-commutator benefit from QITE in the case that the initial state correlation length has an upper bound
\begin{align}
    \xi =  \OO((\frac{1}{2kd}\log N)^\frac{1}{d}).
    \label{eq:correlation-length-bound}
\end{align}

In many physical applications, for example, the initial state is a $\Delta$-gapped ground state that has a constant correlation length $\xi\sim 1/\Delta$. In that case, measuring anti-commutator using QITE is beneficial, especially for sparse device connectivity with a small $d$. This benefit is more considerable for a larger system size $N$. Otherwise, for an initial state with a large correlation length, we provide an alternative method of performing QITE efficiently as discussed in Subsection~\ref{app:Quantum imaginary-time evolution for large correlated states}.

\subsection{Quantum imaginary-time evolution for the SSH model and TIM}\label{eq:qite-ssh-tim}

\begin{figure}
    \centering
    \includegraphics[width=0.6\textwidth]{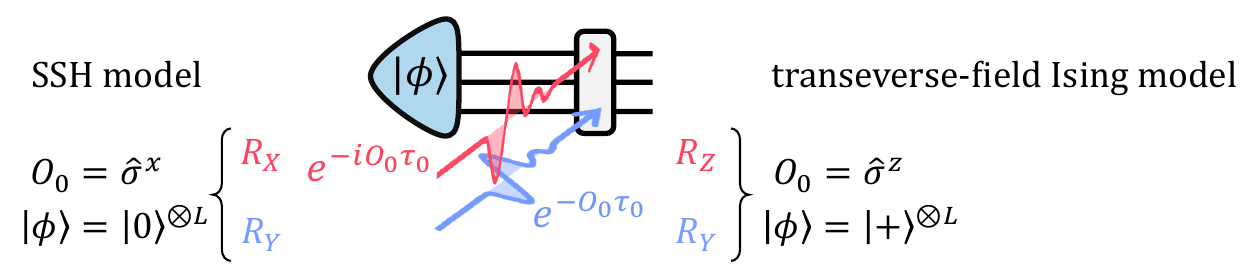}
    \caption{Performing the real-time evolution and quantum imaginary-time evolution (QITE) using orthogonal pulses in the hardware demonstration. The left and right side illustrate the time evolution to measure the $2$-time correlation function of the SSH and the transverse-field Ising model, respectively.}
    \label{fig:single_qubit_qite}
\end{figure}

In our hardware implementation, QITE is used to measure the fermionic spectrum in the Su-Schrieffer-Heeger (SSH) model (See Supplementary Note~\ref{app:fermion-spectrum}) and the real part of OTOC in the transverse-field Ising model (TIM). Their measurement circuits in Fig.~1(\textbf{d}) require QITE of operator $O_0=\sum_{j=0}^{L-1} \cos(kj)\sX_j$ and $O_0=\hat{\sigma}^z_0$, respectively. Because the initial states of the evolution are the product states with zero correlation length, their QITE can be realized efficiently even without any prior measurements. Here we show how the QITE is realized in the hardware implementation.

For the SSH model, the initial state is $\ket{\phi}=\ket{0}^{\otimes L}$. Since $\hat{\sigma}_j^x$ is a single-qubit operator, and the initial state is a product state, the imaginary-time evolution can be realized exactly by single-qubit $R_Y$ rotations
\begin{equation}
    \begin{aligned}
    e^{-\tau  \cos(kj)\hat{\sigma}_j^x}\ket{0}&=e^{-\tau \cos(kj)\hat{\sigma}_j^x}\ket{0}\\
    &=\cosh(\tau\cos(kj))\ket{0}-\sinh(\tau\cos(kj))\ket{1}\\
    &\propto e^{-i\theta_j(k,\tau)\hat{\sigma}_j^y/2}\ket{0},
\end{aligned}
\end{equation}
with the rotation angles $\theta_j(k,\tau)=-2\arctan(\tanh(\tau\cos(kj)))$. Thus, the $R_Y$ rotation is equivalent to QITE of $\hat{\sigma}^x$ for the particular initial state $\ket{0}$.

For TIM, to measure the real part of OTOC, we perform QITE of $\hat{\sigma}_0^z$ on the initial state $\ket{\phi} = \ket{+}^{\otimes L}$. Similar to the case of the SSH model, QITE of $\hat{\sigma}_0^z$ can be realized by the single-qubit $R_Y$ rotation because
\begin{align}
    e^{-\tau \hat{\sigma}_0^z}\ket{+} \propto e^{-i\theta_0(\tau) \hat{\sigma}_0^y/2}\ket{+},
\end{align}
with the rotation angle $\theta_0(\tau)=2\arctan(e^{2\tau})-\pi/2$. 

In these two examples, we see that QITE of single-qubit operators can be realized by the real-time evolution of single-qubit operators at their orthogonal directions (orthogonal in the Pauli basis). This is schematically illustrated by orthogonal pulses in Fig.~1(\textbf{a}) of the main text, and we summarize these two examples in Fig.~\ref{fig:single_qubit_qite}.

\subsection{Quantum imaginary-time evolution by the mid-circuit measurement}\label{app:Quantum imaginary-time evolution for large correlated states}
In Subsection~\ref{app:Run time bounds of the quantum imaginary-time evolution}, we provide the requirement of the initial state correlation length where the ancilla-free measurement using QITE outperforms the Hadamard test. Here we show that this requirement can be circumvented by realizing QITE alternatively using mid-circuit measurement~\cite{Mitarai_19}.

For the Pauli string operator $O=\widetilde{\sigma}$ acting on $k$ qubits, since $\widetilde{\sigma}$ only has eigenvalues $\pm 1$ with two  eigen spaces, it can be rewritten as a linear combination of two projectors $\hat{P}_{\pm}$ onto the two eigen spaces
\begin{align}
    \widetilde{\sigma} = \hat{P}_{\widetilde{\sigma},+}-\hat{P}_{\widetilde{\sigma},-}.
\end{align}
In the parameter-shift rule of QITE, one can choose $\tau=\pm\infty$ such that the QITE is reduced to the projective measurement on $k$ qubits
\begin{align}
    \left\{ \begin{array}{ll}\lim_{\tau\to\infty}\frac{e^{-\tau \widetilde{\sigma}}}{\sqrt{\bra{\psi_0} e^{-2\tau \widetilde{\sigma}}\ket{\psi_0}}}\ket{\psi_0}=
 \frac{\hat{P}_{\widetilde{\sigma},-}}{\sqrt{\bra{\psi_0}\hat{P}_{\widetilde{\sigma},-}\ket{\psi_0}}}\ket{\psi_0}; \\
 \lim_{\tau\to -\infty}\frac{e^{-\tau \widetilde{\sigma}}}{\sqrt{\bra{\psi_0} e^{-2\tau \widetilde{\sigma}}\ket{\psi_0}}}\ket{\psi_0} = \frac{\hat{P}_{\widetilde{\sigma},+}}{\sqrt{\bra{\psi_0}\hat{P}_{\widetilde{\sigma},+}\ket{\psi_0}}}\ket{\psi_0},
  \end{array} \right.
\end{align}
where the two normalized states on the right-hand side represent the states after the projective measurement onto the two eigen spaces. This projective measurement can be realized by sequentially applying (1) Clifford gates transforming $\widetilde{\sigma}$ to a single-qubit $Z_j=\ket{0}\bra{0}_j-\ket{1}\bra{1}_j$; (2) mid-circuit Pauli-$Z_j$ projective measurement onto $\ket{1}_j$ and $\ket{0}_j$ for $\tau\to\infty $ and $\tau\to -\infty $, respectively; (3) the reverse of the Clifford gates. The mid-circuit measurement can be physically realized in current quantum platforms~\cite{PhysRevLett.127.100501,Minev_19}. Equipped with the mid-circuit measurement, QITE can be performed efficiently without requiring a finite correlation length.

\section{Error analysis}
In this note, we estimate the error of calculating the commutator and anti-commutator $c_{\bos{b}}$ using the parameter-shift rule in Eq.~\eqref{eq:c_b-by-parameter-shift-rule}. The error of estimating $c_{\bos{b}}$ comes from four aspects:
\begin{itemize}
    \item The approximate QITE.
    \item Trotter decomposition during the evolution $U(t_i;t_j)$.
    \item Finite number of shots measuring the expectation of $O_{n-1}$.
    \item Noisy quantum circuits.
\end{itemize}
In the following contents, we derive the error bound of $c_{\bos{b}}$ considering the truncation error from QITE, Trotterization of the time evolution, and the statistical error from a finite number of shots. The error from noisy quantum circuits has different effects to $c_{\bos{b}}$ regarding different applications. In the next subsection, we consider its effect in the application of measuring spectrum functions, which is one of the main applications we demonstrate in the main text.
\\\\
\textbf{Truncation error}~---~ The truncation error from QITE has been discussed in Supplementary Note~\ref{app:Run time bounds of the quantum imaginary-time evolution}, where we provide the condition for the existence of the unitary $\hat{U}_{O_j}^{(+)}$ to approximate the imaginary-time evolution $U_{O_j}^{(+)}$
\begin{align}
    \left\| \hat{U}_{O_j}^{(+)}\ket{\psi_0}- U_{O_j}^{(+)}\ket{\psi_0}\right\| \leq \epsilon_{\text{QITE}}.
    \label{eq:QITE-approx}
\end{align}
For convenience, we also denote the physical realization of the real-time evolution $\hat{U}_{O_j}^{(-)} \equiv e^{-i\tau_j O_j}=U_{O_j}^{(-)}$ without truncation errors.

Another source of truncation error is from the Trotter decomposition of the time evolution $U(t_i;t_j)$. If the system Hamiltonian contains non-commuting Pauli strings, its time evolution should be approximated using, e.g., the Trotter-Suzuki decomposition~\cite{SUZUKI1990319}. Specifically, the $p$-order decomposition approximate $U(t_i;t_j)$ using $\hat{U}_p(t_i;t_j)$ with the Trotter error
\begin{align}
    \| \hat{U}_p(t_i;t_j)-U(t_i;t_j)\|=\OO((t_i-t_j)(\delta t)^p\| H\|^{p+1}),
    \label{eq:Trotter-error}
\end{align}
where $\delta t$ is the Trotter step length. Using the $p$-order decomposition, the error of $O_{n-1}$ expectation can be bounded using the triangular inequality
\begin{equation}
    \begin{aligned}
    |\langle O_{n-1}\rangle_{\bos{b},\text{Trot}}-\langle O_{n-1}\rangle_{\bos{b}}|=&|\bra{\phi}\hat{U}^{(b_0)\dagger}_{O_0}\hat{U}_p^{\dagger}(t_1;t_0)\hat{U}^{(b_1)\dagger}_{O_1}\ldots \hat{U}_p^{\dagger}(t_{n-1};t_{n-2})O_{n-1}\hat{U}_p(t_{n-1};t_{n-2})\ldots \hat{U}^{(b_1)}_{O_1}\hat{U}_p(t_1;t_0)\hat{U}^{(b_0)}_{O_0}\ket{\phi}\\
    &-\bra{\phi}U^{(b_0)\dagger}_{O_0}U^{\dagger}(t_1;t_0)U^{(b_1)\dagger}_{O_1}\ldots U^{\dagger}(t_{n-1};t_{n-2})O_{n-1}U(t_{n-1};t_{n-2})\ldots U^{(b_1)}_{O_1}U(t_1;t_0)U^{(b_0)}_{O_0}\ket{\phi}|\\
    =&|\sum_{j=0}^{n-2} (\DD_{L,b_j}+\DD_{L,t_j}+\DD_{R,b_j}+\DD_{R,t_j})|\\
    \leq& \sum_{j=0}^{n-2} (|\DD_{L,b_j}|+|\DD_{L,t_j}|+|\DD_{R,b_j}|+|\DD_{R,t_j}|).
\end{aligned}
\end{equation}
In the second line, we iteratively utilize the decomposition $\hat{A}\hat{B}-AB = (\hat{A}-A)\hat{B}+A(\hat{B}-B)$, for arbitrary matrices $A$ and $B$, and we define
\begin{equation}
    \begin{aligned}
    \DD_{L,b_j}&\equiv \bra{\phi} U^{(b_0)\dagger}_{O_0}\ldots U^{\dagger}(t_{j};t_{j-1})[\hat{U}^{(b_j)\dagger}_{O_j}-U^{(b_j)\dagger}_{O_j}]\hat{U}^{\dagger}_p(t_{j+1};t_j)\ldots O_{n-1}\ldots  \hat{U}^{(b_0)}_{O_0}\ket{\phi};\\
    \DD_{L,t_j}&\equiv \bra{\phi} U^{(b_0)\dagger}_{O_0}\ldots U^{\dagger}(t_{j};t_{j-1})U^{(b_j)\dagger}_{O_j}[\hat{U}^{\dagger}_p(t_{j+1};t_j)-U^{\dagger}(t_{j+1};t_j)]\ldots O_{n-1}\ldots  \hat{U}^{(b_0)}_{O_0}\ket{\phi};\\
    \DD_{R,b_j}&\equiv \bra{\phi} U^{(b_0)\dagger}_{O_0}\ldots O_{n-1}\ldots U(t_{j+1};t_{j})[\hat{U}^{(b_j)}_{O_j}-U^{(b_j)}_{O_j}]\hat{U}_p(t_{j};t_{j-1})\ldots   \hat{U}^{(b_0)}_{O_0}\ket{\phi};\\
    \DD_{R,t_j}&\equiv \bra{\phi} U^{(b_0)\dagger}_{O_0}\ldots O_{n-1}\ldots [\hat{U}_p(t_{j+1};t_j)-U(t_{j+1};t_j)]\hat{U}^{(b_j)}_{O_j}\hat{U}_p(t_{j};t_{j-1})\ldots   \hat{U}^{(b_0)}_{O_0}\ket{\phi}.
\end{aligned}
\end{equation}
Since all the exact or approximate operations are unitaries with the spectral norm $1$, the norms of $\DD_{L(R),b_j}$, $\DD_{L(R),t_j}$ are bounded by
\begin{equation}
    \begin{aligned}
    &|\DD_{L(R),b_j=+}|\leq \epsilon_{\text{QITE}}\|O_{n-1} \|; \quad |\DD_{L(R),b_j=-}|=0;\\
    &|\DD_{L(R),t_j}|\leq \|\hat{U}_p(t_{j+1};t_j)-U(t_{j+1};t_j)\|\|O_{n-1} \|=\OO(\|O_{n-1} \|(t_{j+1}-t_j)(\delta t)^p\|H\|^{p+1}),
\end{aligned}
\end{equation}
where we have used Eq.~(\ref{eq:QITE-approx}, \ref{eq:Trotter-error}) in the first and the second line. Therefore, the error of $O_{n-1}$ expectation is bounded by
\begin{align}
    |\langle O_{n-1}\rangle_{\bos{b},\text{Trot}}-\langle O_{n-1}\rangle_{\bos{b}}| \leq \left[n_{+,\bos{b}} \epsilon_{\text{QITE}}+\OO((t_{n-1}-t_0)(\delta t)^p\|H\|^{p+1})\right] \|O_{n-1} \|,
\end{align}
where $n_{+,\bos{b}}$ is the number of $+$ components in the tuple $\bos{b}$.
\\\\
\textbf{Statistical error}~---~ The measurement process introduces the statistical error due to a finite number of shots $N_{\text{shot}}$. Assuming that the repeated shots are independent and identically distributed events, the statistical error can be estimated from the variance of the measured observable $O_{n-1}$~\cite{lin2022lecturenotesquantumalgorithms}
\begin{align}
    |\langle O_{n-1}\rangle_{\bos{b},\text{meas}}-\langle O_{n-1}\rangle_{\bos{b},\text{Trot}}| = \sqrt{\frac{\Var(O_{n-1})}{N_{\text{shot}}}}=\sqrt{\frac{\langle O_{n-1}^2\rangle_{\bos{b}}-\langle O_{n-1}\rangle_{\bos{b}}^2}{N_{\text{shot}}}}\leq \frac{\| O_{n-1}\|}{\sqrt{N_{\text{shot}}}},
\end{align}
where the second equality comes from the definition of the variance, and the third inequality follows from the realness of the $O_{n-1}$ expectation, such that $\langle O_{n-1}\rangle_{\bos{b}}^2\geq 0$.
\\\\
\textbf{Combined error}~---~ Combining the truncation and the statistical errors, the total error of the expectation is bounded by
\begin{equation}
    \begin{aligned}
    |\langle O_{n-1}\rangle_{\bos{b},\text{meas}}-\langle O_{n-1}\rangle_{\bos{b}}| &\leq |\langle O_{n-1}\rangle_{\bos{b},\text{meas}}-\langle O_{n-1}\rangle_{\bos{b},\text{Trot}}|+|\langle O_{n-1}\rangle_{\bos{b},\text{Trot}}-\langle O_{n-1}\rangle_{\bos{b}}|\\
    &\leq \|O_{n-1} \|\left[\frac{1}{\sqrt{N_{\text{shot}}}}+n_{+,\bos{b}} \epsilon_{\text{QITE}}+\OO((t_{n-1}-t_0)(\delta t)^p\|H\|^{p+1})\right].
\end{aligned}
\end{equation}
Taking this bound to Eq.~\eqref{eq:c_b-by-parameter-shift-rule}, the error of estimating $c_{\bos{b}}$ is bounded by
\begin{equation}
    \begin{aligned}
    |c_{\bos{b},\text{meas}}-c_{\bos{b}}|&\leq \sum_{\bos{\xi}\in\{0,1\}^{n-1}} |\langle O_{n-1}\rangle_{\bos{b}}(\bos{\tau}_{\bos{\xi}})_{\text{meas}}-\langle O_{n-1}\rangle_{\bos{b}}(\bos{\tau}_{\bos{\xi}})|\\
    &\leq 2^{n-1} \|O_{n-1} \|\left[\frac{1}{\sqrt{N_{\text{shot}}}}+n_{+,\bos{b}} \epsilon_{\text{QITE}}+\OO((t_{n-1}-t_0)(\delta t)^p\|H\|^{p+1})\right] ,
\end{aligned}
\end{equation}
where we choose the real(iimaginary) evolution time $\tau_-=\pi/4$ ($\tau_+=\frac{1}{2}\log(1+\sqrt{2})$) in the $(n-1)$-tuple $\bos{\tau}_{\bos{\xi}} \equiv ((-1)^{\xi_0}\tau_{b_0},\ldots,(-1)^{\xi_{n-2}}\tau_{b_{n-2}})$. Typically, the operator $O_{n-1}$ is intensive, or localized in either space or momentum domain, such that the norm $\| O_{n-1}\|$ is constant to the system size $N$, and also a constant $n\sim\OO(1)$ is enough for physically relevant applications. Thus, to keep the error of $c_{\bos{b}}$ smaller than a constant, $N_{\text{shot}}$ and $\epsilon_{\text{QITE}}$ can be constant to the system size $N$, and $\delta t$ should decrease polynomially with respect to $N$ to suppress the polynomially increased Hamiltonian norm $\| H\|$. Thus, the total gate counts are polynomially increased with respect to the system size, ensuring the scalability of our measurement method.

\subsection{Circuit noise effect on spectrum functions}\label{app:Circuit noise effect on spectrum functions}
Spectrum functions of a quantum system can be measured by the Fourier transformation of 2-time correlation functions. Assume that $\ket{E_0}$ is the ground state of the quantum system. We define the 2-time correlation functions
\begin{align}
   C_O(t)\equiv \bra{E_0}O(t)O(0)^{\dagger} \ket{E_0},
\end{align}
where $O^{(\dagger)}$ is the annihilation (creation) operator of some specific particles pocessing a given quantum number. The Fourier transformation of $C_O(t)$ reads
\begin{align}
    \widetilde{C}_O(\omega) = \int_{-\infty}^{\infty} \ud t~ e^{-i\omega t} C_O(t) = \sum_n |\bra{E_n}O\ket{E_0}|^2 2\pi \delta (\omega -(E_n-E_0)),
    \label{eq:C-frequency-noiseless}
\end{align}
which is a summation of $\delta$-functions considering infinite evolution time.

\begin{figure}
    \centering
    \includegraphics[width=0.5\textwidth]{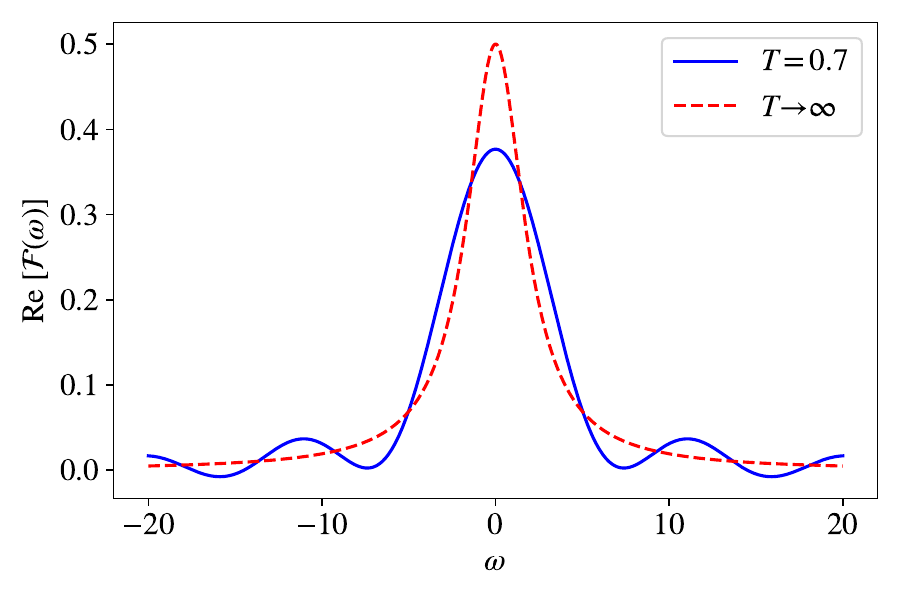}
    \caption{The real part of the smearing function $\mathcal{F}(\omega)$ with a finite $T$ and $T\to \infty$, denoted by blue solid and red dashed lines, respectively. $\text{Re }\mathcal{F}(\omega)$ is peaked at $\omega=0$. Here we set $\Gamma=5$. The $T\to \infty$ curve has the width of half maximum $\Delta\omega = 2\Gamma$.}
    \label{fig:smearing-function}
\end{figure}

Noise in the quantum circuit and a finite evolution time $T$ deform $\widetilde{C}_O(\omega)$. As an example, we assume that the physical error in the quantum device after applying one layer of the Trotter circuit is a global depolarizing channel with a depolarizing probability $p$. The depolarizing channel is a good approximation of quantum noise in realistic quantum devices performing Trotterized evolutions~\cite{PhysRevLett.127.270502,PhysRevResearch.7.023032}. The depolarizing noise lead to an exponential decay of $C_O(t)$ as a function of time. Thus, the noisy $C_O(t)$ evaluated within a finite evolution time $t\in[0,T]$ is derived by multiplying $C_O(t)$ with a window function:
\begin{align}
    C_O'(t) = e^{-\Gamma t} [\theta(t)-\theta(t-T)] C_{O}(t),
    \label{eq:C_O'(t)}
\end{align}
where $\Gamma=-\log(1-p)/\delta t\approx p/\delta t$ is the decay rate of the Trotterized evolution with the Trotter step length $\delta t$. Then, the Fourier transformation of $C_O'(t)$ can be derived according to the convolution theorem
\begin{equation}
    \begin{aligned}
    \widetilde{C}_O'(\omega) &=\frac{1}{2\pi} \int_{-\infty}^{\infty} \ud\nu~ \mathcal{F}(\omega-\nu) \widetilde{C}_O(\nu)\\
    &= \sum_n |\bra{E_n}O\ket{E_0}|^2 \mathcal{F}(\omega-(E_n-E_0)).
\end{aligned}
\end{equation}
The second line is derived by taking Eq.~\eqref{eq:C-frequency-noiseless} to $\widetilde{C}_O(\nu)$, and $\mathcal{F}(\omega)$ is the Fourier transformation of the window function in Eq.~\eqref{eq:C_O'(t)}
\begin{align}
    \mathcal{F}(\omega) = \int_{-\infty}^{\infty} \ud t~ e^{-i\omega t} e^{-\Gamma t} [\theta(t)-\theta(t-T)] = \int_{0}^{T} \ud t~ e^{-(i\omega+\Gamma) t} = \frac{1-e^{-(\Gamma+i\omega)T}}{\Gamma+i\omega}.
\end{align}
The real part of $\mathcal{F}(\omega)$ with a finite $T$ and $T\to \infty$ is shown in Fig.~\ref{fig:smearing-function}. In the limit $T\to \infty$, $\text{Re }\mathcal{F}(\omega)$ is peaked at $\omega=0$ and has the width of half maximum $\Delta\omega = 2\Gamma$. Thus, comparing the noisy spectrum function $\widetilde{C}_O'(\omega)$ with the noiseless one in Eq.~\eqref{eq:C-frequency-noiseless}, the noisy quantum circuit deform the $\delta$-function in $\widetilde{C}_O(\omega)$ to the smeared one $\mathcal{F}(\omega-(E_n-E_0))$ peaked at the same point $\omega = E_n-E_0$.  Therefore, the depolarizing noise do not shift the peak positions of the spectrum function, but reduce the resolution of spectral lines. Specifically, two spectral lines at $E_n-E_0$ and $E_m-E_0$ can be distinguished only if
\begin{align}
    |E_n-E_m|\gtrsim 2\Gamma\approx 2p/\delta t.
\end{align}
We see that the spectral resolution decreases for stronger circuit noise $p$ and smaller Trotter step lengths $\delta t$. For a finite evolution time $T$, $\mathcal{F}(\omega)$ has side lobes that may be erroneously identified as spectral lines. The misidentification is more severe for smaller $T$ and stronger circuit noise $p$.

\section{Fermionic spectrum of the Su-Schrieffer-Heeger model}\label{app:fermion-spectrum}

\begin{figure*}
    \centering
    \includegraphics[width=0.98\textwidth]{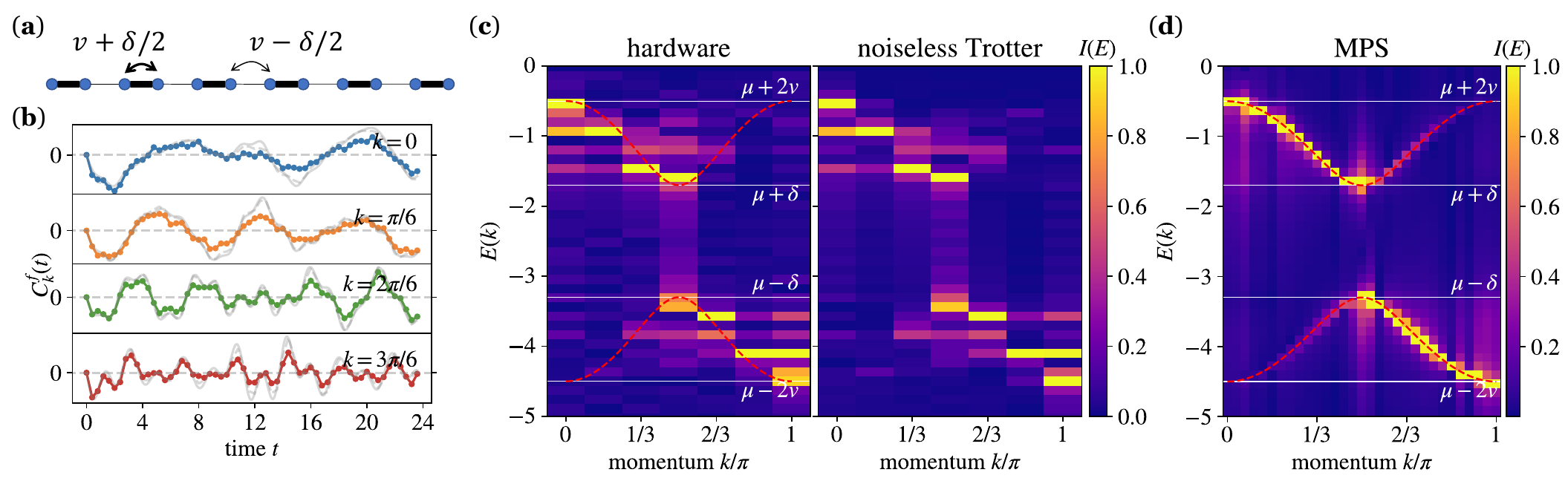}
    \cprotect\caption{Measurement of the fermionic spectrum of the Su-Schrieffer-Heeger (SSH) model (\textbf{a}) Lattice and fermionic hopping of the SSH model. (\textbf{b}) Measurement results of 2-time correlation function $C^f_k(t)$ on \verb|ibm_torino| versus time $t$ with four different momenta $k$. The grey solid and dashed lines denote exact diagonalization and noiseless Trotter results. (\textbf{c}) The normalized fermionic spectrum of the SSH model derived by the Fourier transformation of curves in (\textbf{b}). The red dashed curves denote the analytic spectrum. (\textbf{d}) Simulation results using matrix product state (MPS) on an $L=60$ lattice.}
    \label{fig:ssh-result}
\end{figure*}

We apply our method to measure the fermionic spectrum of the one-dimensional Su-Schrieffer-Heeger (SSH) model, which describes the staggered hopping interaction as in polyacetylene with a topologically non-trivial phase~\cite{Girvin_Yang_2019}. With the open boundary condition (OBC), the Hamiltonian reads
$
    H_{\text{SSH}} = -\sum_{j=0}^{L-2}(v+(-1)^j\delta/2)(\hat{c}^{\dagger}_j\hat{c}_{j+1}+h.c.)-\mu \sum_{j=0}^{L-1}\hat{c}^{\dagger}_j\hat{c}_j$,
where the first and second terms are the staggered hopping (as illustrated in Fig.~\ref{fig:ssh-result}(\textbf{a})) and the on-site energy with chemical potential $\mu$, respectively. We measure the real part of the fermionic 2-time correlation function
\begin{equation}
    \begin{aligned}
     C_k^{f}(t) &\equiv \operatorname{Re}(-i \bra{\Omega}[\tilde{c}_k(t),\tilde{c}^{\dagger}_k(0)]_+\ket{\Omega}),
    \label{eq:fermionic-two-time-correlation-function}
\end{aligned}
\end{equation}
where $\tilde{c}^{(\dagger)}_k$ denotes momentum definite fermionic annihilation (creation) operator with a lattice momentum $k=2\pi n/L, n\in[-L/2, L/2]$. In the hardware implementation, we choose a large chemical potential $\mu$ such that the ground state is $\ket{\Omega}=\ket{00\ldots 0}$. In Supplementary Note~\ref{app:Models description and numerical results}, we show that $C_k^{f}(t)$ can be evaluated by taking $O_0= \sum_{j=0}^{L-1} \cos(kj)\sX_j$ and $O_1=\sY_0$ in the measurement circuit Fig.~1(\textbf{d}) of the main text. Additionally, the imaginary-time evolution of $O_0$ on $\ket{\Omega}$ can be performed using single-qubit $R_{Y}$ rotations as detailed in Supplementary Note~\ref{app:high-order-qite}.

For the $L=12$ SSH model, the hardware measurement results of $C_k^{f}(t)$ with different evolution time $t$ are shown in Fig.~\ref{fig:ssh-result}(\textbf{b}). Here, the Trotter evolution circuits of the SSH model are compressed using the algebraic compression method~\cite{PhysRevA.105.032420}. The Fourier transformation of $C_k^{f}(t)$ gives the normalized fermionic spectrum shown in Fig.~\ref{fig:ssh-result}(\textbf{c}). The positions of energy peaks are in agreement with the noiseless Trotter results. In this demonstration, with the staggered hopping parameter $\delta=0.8$ in $H_{\text{SSH}}$, the spectrum possesses a gap of $2\delta$ at $k=\pi/2$. The energy peaks slightly deviate from the analytic ones (red dashed curves in Fig.~\ref{fig:ssh-result}(\textbf{c})) due to the finite volume effect. To see this, Fig.~\ref{fig:ssh-result}(\textbf{d}) shows the simulation results with $L=60$ using the matrix product state (MPS) method. The spectrum agrees with the analytic one in this larger-volume simulation.

\section{Model description and the correlation function}\label{app:Models description and numerical results}

\subsection{Schwinger model}
In axial gauge with open boundary conditions (OBCs), using the Jordan-Wigner transformation, the Hamiltonian of the massive Schwinger model reads
\begin{equation}
\begin{aligned}
    H_{\text{SW}} = H_{\text{m}}+ H_{\text{kin}}+ H_{\text{ele}} = \frac{m}{2}\sum_{j=0}^{2L-1}(-1)^j\hat{\sigma}^z_j+\frac{1}{2}\sum_{j=0}^{2L-2}(\hat{\sigma}_j^+\hat{\sigma}_{j+1}^-+h.c.)+\frac{g^2}{2}\sum_{j=0}^{2L-2}(\sum_{l\leq j}\hat{Q}_l)^2,
    \label{eq:Schwinger-Hamiltonian}
\end{aligned}
\end{equation}
where $\hat{Q}_l=(\hat{\sigma}^z_l+(-1)^l)/2$ is the charge operator. We choose the bare mass $m=0.5$ and coupling $g=0.3$ in the hardware implementation. The hadron spectrum can be obtained by the Fourier transformation of the hadron Green's function
\begin{align}
    C_k^b(t)\equiv -i\bra{\Omega}[\tilde{h}_k(t), \tilde{h}_k(0)]_-\ket{\Omega},
\end{align}
where $\ket{\Omega}$ is the ground state of the Schwinger model. $\tilde{h}_k=\sum_i e^{-ikj}\hat{h}_j$ with a lattice momentum $k=2\pi n/L,n\in[-L/2,L/2]$, which is the Fourier transform of the hadron operator 
\begin{align}
    \hat{h}_j=-i(\hat{\sigma}^+_{2j}\hat{\sigma}^-_{2j+1}+h.c.)
    \label{eq:h_j_spin}
\end{align}
obtained by the Jordan-Wigner transformation of $\bar{\psi}_j\gamma^5\psi_j$.

To reduce the quantum computing resource needs, we apply some simplifications in the hardware implementation. First,  we approximate the ground state $\ket{\Omega}$ by the bare vacuum $\ket{\Omega_0}=\ket{01\ldots 01}$, i.e., the ground state of the mass term $m\sum_{j=0}^{2L-1}(-1)^j\hat{\sigma}^z_j/2$ with $m>0$. Thus, in the practical implementation, we calculate the approximation of $C_k^b(t)$ 
\begin{align}
    {C}^b_{0,k}(t)\equiv -i\bra{\Omega_0}[\tilde{h}_k(t), \tilde{h}_k(0)]_-\ket{\Omega_0}.
\end{align}
This approximation is feasible because the spectrum of the hadron $\tilde{h}_k$ is the minimum excitation energy of the Hamiltonian with a given momentum $k$. Using this approximation, the minimum excitation energy can still be identified, as long as the overlap $|\langle\Omega_0|\Omega\rangle|$ is not very small ($|\langle\Omega_0|\Omega\rangle|=0.577$ in the hardware implementation according to the exact diagonalization). Second, the Hamiltonian $H_{\text{SW}}$ contains the long-range interactions in the electric energy $H_{\text{ele}}$ of Eq.~\eqref{eq:Schwinger-Hamiltonian}. The long-range interaction challenges the quantum simulation of the Schwinger model on superconducting quantum chips with finite qubit connectivity. Thus, we perform the simulation using a simplified Hamiltonian $H_{\text{SW}}^{\text{tr}}$ described in Ref.~\cite{PhysRevD.109.114510}. It truncates the long-range interactions that are spatially far apart. The truncated Hamiltonian $H_{\text{SW}}^{\text{tr}} \equiv H_{\text{m}}+ H_{\text{kin}}+ H_{\text{ele}}^{\text{tr}}$ has the electric energy
\begin{equation}
    \begin{aligned}
    H_{\text{ele}}^{\text{tr}}=&\frac{g^2}{2}\left\{ \sum_{j=0}^{L/2-1}\Big[(\frac{L}{2}-\frac{3}{4}-j)\sZ_{2j}\sZ_{2j+1}+(j+\frac{1}{4})\sZ_{L+2j}\sZ_{L+2j+1}\Big]\right.\\
    &+\frac{1}{2}\sum_{j=1}^{L/2-2}(2\sZ_{2j}+\sZ_{2j+1}-\sZ_{L+2j}-2\sZ_{L+2j+1}) +\frac{1}{2}(2\sZ_0+\sZ_1+\sZ_{L-2}-\sZ_{L+1}-\sZ_{2L-2}-2\sZ_{2L-1})\\
    &+\sum_{j=0}^{L/2-2}\left[ (\frac{L}{2}-\frac{5}{4}-j)(\sZ_{2j}+\sZ_{2j+1})\sZ_{2j+2}+(\frac{L}{2}-\frac{7}{4}-j)(\sZ_{2j}+\sZ_{2j+1})\sZ_{2j+3}\right. \\
    &+\left.\left. (j+\frac{1}{4})(\sZ_{L+2j+2}+\sZ_{L+2j+3})\sZ_{L+2j}+(j+\frac{3}{4})(\sZ_{L+2j+2}+\sZ_{L+2j+3})\sZ_{L+2j+1}\right]\right\},
\end{aligned}
\end{equation}
which contains up to the next-next nearest-neighbor interaction $\sZ_j\sZ_{j+3}$. The truncated Hamiltonian $H_{\text{SW}}^{\text{tr}}$ preserves the spatial reflection symmetry. It is a good approximation to the exact one $H_{\text{SW}}$, as long as the initial state $\ket{\Omega_0}$ is short-range correlated~\cite{PhysRevD.109.114510}. Third, since the Schwinger model has momentum as a good quantum number in the infinite volume limit,  $C^b_{0,k}(t)$ can be evaluated by the momentum projection  
\begin{equation}
    \begin{aligned}
    C^b_{0,k}(t)
    &=-i\bra{\Omega_0}[\tilde{h}_k(t), \sum_{k'} e^{ik'j_0}\tilde{h}_{k'}(0)]_-\ket{\Omega_0}/e^{ikj_0}\\
    &=-i\sqrt{L}\bra{\Omega_0}[\tilde{h}_k(t), \hat{h}_{j_0}(0)]_-\ket{\Omega_0}/e^{ikj_0}\\
    &=-i\sum_j e^{-ik(j+j_0)}\bra{\Omega_0}[\hat{h}_j(t),\hat{h}_{j_0}(0)]_-\ket{\Omega_0},
    \label{eq:cb0k-expression}
\end{aligned}
\end{equation}
where $j_0$ is an arbitrary spatial site. In practice, to minimize the boundary effects, we choose $j_0=\lceil L/2\rceil$ to be at the center of the one-dimensional chain. Thus, the hadron spectrum can be obtained from the Fourier transformation of $C^b_{0,k}(t)$.

In the numerical results of the hadron spectrum, the signal $C^b_{0,k}(t)$ is doubled in the negative time direction according to the relation
\begin{align}
    C^b_{0,k}(t) =- C^b_{0,k}(-t).
\end{align}
Because the truncated Schwinger model $H_{\text{SW}}^{\text{tr}}$ has the time-reversal symmetry, i.e., $H_{\text{SW}}^{\text{tr}}= H_{\text{SW}}^{\text{tr}*}$, where $(\cdot)^*$ denotes the complex conjugation for each element in the matrix representation of $H_{\text{SW}}^{\text{tr}}$. To see this, we first show that $C^b_{0,k}(t)$ is a real function. Taking the Hermitian conjugation of Eq.~\eqref{eq:cb0k-expression} gives
\begin{align}
    (C^b_{0,k}(t))^{\dagger} = i\sum_j e^{ik(j+j_0)}\bra{\Omega_0} (\hat{h}_{j_0})^{\dagger} e^{iH_{\text{SW}}^{\text{tr}}t}(\hat{h}_{j})^{\dagger}e^{-iH_{\text{SW}}^{\text{tr}}t}-e^{iH_{\text{SW}}^{\text{tr}}t} (\hat{h}_{j})^{\dagger}e^{-iH_{\text{SW}}^{\text{tr}}t}(\hat{h}_{j_0})^{\dagger}\ket{\Omega_0}.
\end{align}
Since $\hat{h}_j$ is anti-Hermitian $(\hat{h}_j)^{\dagger} = -\hat{h}_j$ according to Eq.~\eqref{eq:h_j_spin}, we have
\begin{align}
    (C^b_{0,k}(t))^{\dagger} = -i\sum_j e^{ik(j+j_0)}\bra{\Omega_0}[\hat{h}_j(t), \hat{h}_{j_0}(0)]_-\ket{\Omega_0} = C^b_{0,-k}(t) = C^b_{0,k}(t),
\end{align}
where the last equality holds due to the spatial reflection symmetry of $H_{\text{SW}}^{\text{tr}}$.

On the other hand, taking the complex conjugation for each multiplication element of $C^b_{0,k}(t)$ leads to
\begin{equation}
    \begin{aligned}
    C^b_{0,k}(t) = (C^b_{0,k}(t))^{*} &= (-i)^*\sum_j e^{(-i)^*k(j+j_0)}\bra{\Omega_0}^*[e^{(i)^*H_{\text{SW}}^{\text{tr}*}t}(\hat{h}_j)^*e^{(-i)^*H_{\text{SW}}^{\text{tr}*}t},(\hat{h}_{j_0})^*]_-\ket{\Omega_0}^*\\
    &= i\sum_j e^{ik(j+j_0)}\bra{\Omega_0}[e^{-iH_{\text{SW}}^{\text{tr}}t}(-\hat{h}_j)e^{iH_{\text{SW}}^{\text{tr}}t},(-\hat{h}_{j_0})]_-\ket{\Omega_0}\\
    &=-C_{0,-k}^b(-t)\\
    &=-C_{0,k}^b(-t).
\end{aligned}
\end{equation}
Therefore, we can double the signal $C^b_{0,k}(t)$ in the negative time direction.

\subsection{Systematic error in the hadron spectrum}

\begin{figure}
    \centering
    \includegraphics[width=0.90\textwidth]{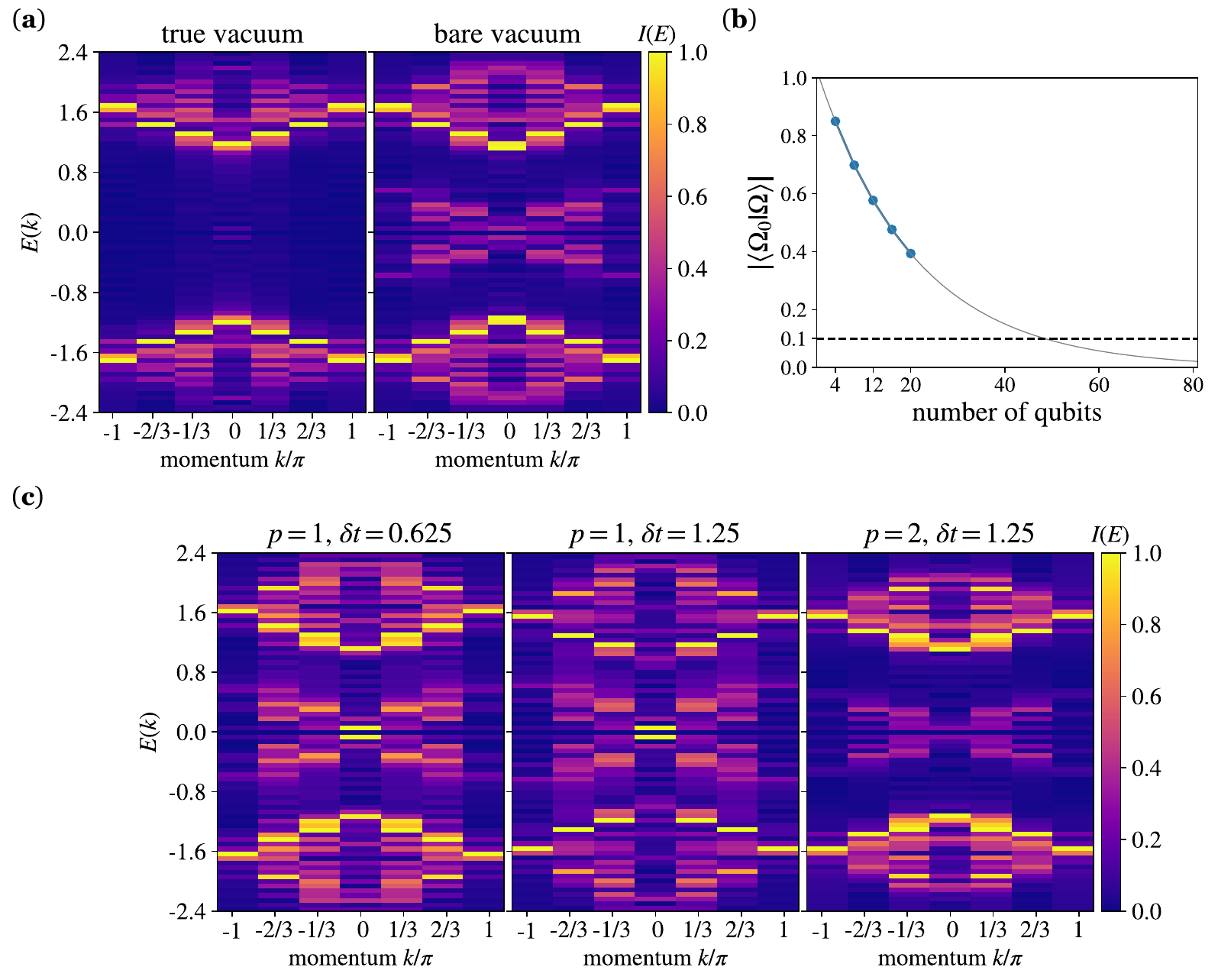}
    \caption{Effect of the approximate ground state and the Trotter decomposition to the hadron spectrum using the classical simulator. (\textbf{a}) Comparison of the evaluated hadron spectrum using the true vacuum and the bare vacuum as the initial state. (\textbf{b}) The overlap between the true vacuum $\ket{\Omega}$ and the bare vacuum $\ket{\Omega_0}$ decreases as the number of qubits increases. The gray solid line is the fitting result using an exponential decay curve. (\textbf{c}) Normalized hadron spectrum derived by $p$-order Trotter decomposition with Trotter step length $\delta t$. The second-order Trotter decomposition performs best with the minimum low-frequency noise due to the Trotter error.}
    \label{fig:hadron-spectrum-app}
\end{figure}

In the measurement scheme of the hadron spectrum, the systematic error comes from three aspects
\begin{itemize}
    \item The true vacuum $\ket{\Omega}$ is approximated by the bare vacuum $\ket{\Omega_0}$.
    \item The Hamiltonian evolution is approximated using Trotter decomposition.
    \item Noise in the quantum circuits.
\end{itemize}
In this subsection, we first discuss the impact of the first two origins by numerical simulations. The numerical simulation is performed on the Schwinger model with the lattice size $L=6$ ($12$ qubits) and Hamiltonian parameters $m=0.5$, $g=0.3$, which are the same as the one in the hardware implementation of the main text.

Fig.~\ref{fig:hadron-spectrum-app}(\textbf{a}) shows the normalized hadron spectra obtained by exact diagonalization with the initial state $\ket{\phi}$ to be the true vacuum $\ket{\Omega}$ (the ground state of $H_{\text{SW}}^{\text{tr}}$) and the bare vacuum $\ket{\Omega_0}$ (the ground state of the mass term $H_{\text{m}}$). We see that in the current lattice and parameters setup, the position of energy peaks in the bare vacuum is consistent with the one in the true vacuum. The bare vacuum results in low- and high-energy spectral lines. But these spectral lines are distinguished from the signal. Thus the hadron spectrum with the bare vacuum well approximates the one with the true vacuum.

The substitution of the true ground state with the bare vacuum state allows us to bypass the quantum circuit overhead associated with ground state preparation. This simplification is feasible in the condition that the bare vacuum and the true vacuum have a large overlap $|\langle{\Omega_0}|\Omega\rangle|$. Different Hamiltonian parameters and lattice sizes lead to different overlaps. For example, the overlap $|\langle{\Omega_0}|\Omega\rangle|$ would decrease as the lattice size increases. In Fig.~\ref{fig:hadron-spectrum-app}(\textbf{b}), we numerically calculate the overlap $|\langle{\Omega_0}|\Omega\rangle|$ versus the number of qubits in the lattice. The overlap is decreased as the number of qubits increases, and we extrapolate the curve using an exponential decay function. According to the extrapolation, the overlap is smaller than $0.1$ as the number of qubits larger than around $50$, such that the spectrum of the hadron excited from the true ground state is hard to be obtained. Thus, to evaluate the hadron spectrum on a larger lattice, we need the ground state preparation circuit to ensure that the hadron spectrum is distinguishable from the low- and high-energy spectral lines. For example, one can use quantum circuits given by variational quantum eigensolver (VQE)~\cite{PhysRevD.109.114510} or the adiabatic process~\cite{Ghim:2024pxe}.
\begin{figure}
    \centering
    \includegraphics[width=0.65\textwidth]{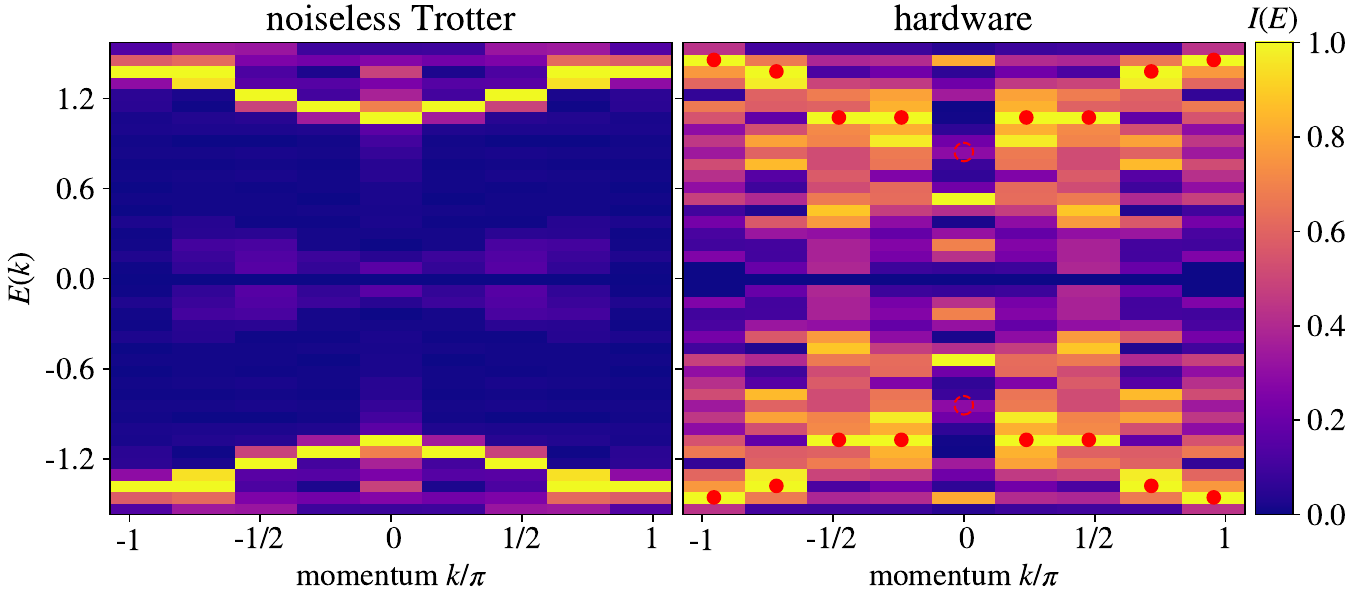}
    \cprotect\caption{Normalized hadron spectrum measured on \verb|ibm_torino| using $16$ qubits. The red dots in the hardware results highlight the energy peaks in the derived spectra for each lattice momentum $k$. Compared with the noiseless Trotter results, the low-energy spectral lines led by noise in the quantum circuits are significant.}
    \label{fig:16-qubit-results}
\end{figure}
Using the bare vacuum as the initial state, we perform the Hamiltonian evolution using the first-order ($p=1$) and the second-order ($p=2$) Trotter decomposition. The resulting normalized hadron spectra are shown in Fig.~\ref{fig:hadron-spectrum-app}(\textbf{c}). Comparing the $p=1$ results with Trotter step length $\delta t=0.625$ and $\delta t=1.25$, the low energy spectral lines are more significant for $\delta t=1.25$ especially at $k=0$, such that the hadron spectral line at $k=0$ is all most invisible. Additionally, comparing the position of energy peaks in $\delta t=0.625$ and $\delta t=1.25$, the energy is smaller for larger Trotter step length, and their energy values are both smaller than the exact diagonalization values in Fig.~\ref{fig:hadron-spectrum-app}(\textbf{a}). Thus, the Trotter error makes the estimated spectrum smaller, and it is the main reason that the measured hadron mass $m_hc^2_{\text{CA+filter}}=1.119(36)$ in our hardware implementation is smaller than the exact $m_hc^2|_{\text{ex}} = 1.162$. 

The signals are improved by the $p=2$ Trotter decomposition compared to the ones of $p=1$. The number of quantum gates in one Trotter step of $p=2$ decomposition is no more than that in two steps of $p=1$ decomposition~\cite{SUZUKI1990319}. Thus, the number of quantum gates used in Fig.~\ref{fig:hadron-spectrum-app}(\textbf{c}) with $p=2,\delta t=1.25$ is smaller than that in Fig.~\ref{fig:hadron-spectrum-app}(\textbf{c}) with $p=1,\delta t=0.625$. But the former exhibits a cleaner signal than the latter. For this reason, we use second-order Trotter decomposition in the hardware implementation.

To demonstrate the impact of noise in quantum circuits, we implement our algorithm on a $L=8$ lattice with $16$ qubits. The resulting hadron spectrum is shown in Fig.~\ref{fig:16-qubit-results}. Compared with the noiseless Trotter results, the low-energy spectral lines resulting from noise in the quantum circuits are significant. The possible hadron spectral line at $k=0$ labeled by the dashed circle is hard to distinguish from other lines in the same column, partially due to the finite evolution time and large decoherence as discussed in Note~\ref{app:Circuit noise effect on spectrum functions}. The derived hadron spectrum of the Schwinger model at size $L=8$ has a larger systematic error than $L=6$ due to noise in the superconducting quantum chip. Here, we do not adopt the classical signal-processing method, and the 2-time correlation function is measured only once. The noiseless hadron spectrum could be reproduced using more advanced quantum devices and by adopting the classical signal-processing method, as well as other conventional error mitigation techniques.

\subsection{Su-Schrieffer-Heeger model}

With open boundary conditions (OBCs), using the Jordan-Wigner transformation, the Hamiltonian of the Su-Schrieffer-Heeger (SSH) model has a spin representation
\begin{equation}
    \begin{aligned}
    H_{\text{SSH}} = &-\sum_{j=0}^{L-2}(v+(-1)^j\delta/2)(\hat{\sigma}_j^+\hat{\sigma}_{j+1}^-+h.c.)+\frac{\mu}{2} \sum_{j=0}^{L-1}\sZ_j.
    \label{eq:SSH-Pauli}
\end{aligned}
\end{equation}
We choose the hopping interaction $v=1$, $\delta=0.8$, and the on-site energy $\mu=-2.5$ in the hardware demonstration. The 2-time anti-commutator $C^f_k(t)$ is measured
with the initial state $\ket{\Omega}=\ket{\bar{0}}\equiv\ket{00\ldots 0}$ by the following quantity
\begin{equation}
    \begin{aligned}
     \check{C}^f_k(t) \equiv \frac{1}{2}\bra{\bar{0}}[\sY_0(t),\sum_j \cos(kj)\sX_j]_+\ket{\bar{0}}.
    \label{eq:fermionic-two-time-correlation-function-app}
\end{aligned}
\end{equation}
In the following content, we show that in the infinite volume limit, $\check{C}^f_k(t)$ equals to $C^f_k(t)$ defined in Eq.~\eqref{eq:fermionic-two-time-correlation-function} of the main text.

Using the Jordan-Wigner transformation, the spin operator in Eq.~\eqref{eq:fermionic-two-time-correlation-function-app} can be transformed into fermionic operators
\begin{align}
    \check{C}^f_k(t) = \frac{1}{2}\bra{\bar{0}}[i(\hat{c}^{\dagger}_0(t)-\hat{c}_0(t)),\sum_j \cos(kj) (\hat{c}^{\dagger}_j+\hat{c}_j)]_+\ket{\bar{0}}.
\end{align}
Here, all the parity of fermionic operators is trivial since the initial state $\ket{\bar{0}}$ corresponds to zero fermion. Then, taking the Fourier transformation of these spatial definite operators gives
\begin{equation}
\begin{aligned}
    \check{C}^f_k(t) &= \frac{1}{2}\bra{\bar{0}}[i\sum_{k'}(\tilde{c}^{\dagger}_{k'}(t)-\tilde{c}_{k'}(t)),\frac{1}{2} (\tilde{c}_k^{\dagger}+\tilde{c}_{-k}^{\dagger}+\tilde{c}_k+\tilde{c}_{-k})]_+\ket{\bar{0}}\\
    &= \frac{i}{4} (\bra{\bar{0}} [\tilde{c}_k^{\dagger}(t), \tilde{c}_k]_+\ket{\bar{0}}+\bra{\bar{0}} [\tilde{c}_{-k}^{\dagger}(t), \tilde{c}_{-k}]_+\ket{\bar{0}}-\bra{\bar{0}} [\tilde{c}_k(t), \tilde{c}^{\dagger}_k]_+\ket{\bar{0}}-\bra{\bar{0}} [\tilde{c}_{-k}(t), \tilde{c}_{-k}^{\dagger}]_+\ket{\bar{0}}\\
    &= \operatorname{Re}(-i \bra{\bar{0}} [\tilde{c}_k(t), \tilde{c}^{\dagger}_k]_+\ket{\bar{0}})\\
    &= C^f_k(t).
    \label{eq:Cfk(t)}
\end{aligned}
\end{equation}
In the second line, we assume that the momentum is a good quantum number, such that $\bra{\bar{0}} [\tilde{c}_{k'}^{\dagger}(t), \tilde{c}_k]_+\ket{\bar{0}}=\bra{\bar{0}} [\tilde{c}_{k'}(t), \tilde{c}^{\dagger}_k]_+\ket{\bar{0}}=0$ for $k'\neq k$. It holds exactly for $H_{\text{SSH}}$ in the infinite volume limit. In the third line, we use the property that the SSH model is symmetric under spatial reflection so that $\bra{\bar{0}} [\tilde{c}_k^{\dagger}(t), \tilde{c}_k]_+\ket{\bar{0}}=\bra{\bar{0}} [\tilde{c}_{-k}^{\dagger}(t), \tilde{c}_{-k}]_+\ket{\bar{0}}$, and the relation $i\bra{\bar{0}} [\tilde{c}_k^{\dagger}(t), \tilde{c}_k]_+\ket{\bar{0}} = (-i\bra{\bar{0}} [\tilde{c}_k(t), \tilde{c}^{\dagger}_k]_+\ket{\bar{0}})^{\dagger}$. Thus, $\check{C}^f_k(t)$ equals to $C^f_k(t)$ in the infinite volume limit. One can estimate the fermionic spectrum using the anti-commutator in Eq.~\eqref{eq:fermionic-two-time-correlation-function-app}.

Similar to the Schwinger model, in the numerical results of the SSH model, the signal $\check{C}^f_k(t)$ is doubled in the negative time direction according to
\begin{align}
    \check{C}^f_k(t) = -\check{C}^f_k(-t).
\end{align}
due to the time-reversal symmetry of the SSH model. According to Eq.~\eqref{eq:Cfk(t)}, $\check{C}^f_k(t)$ is a real function. Thus, taking complex conjugation to each multiplication element in $\check{C}^f_k(t)$ gives 
\begin{equation}
    \begin{aligned}
    \check{C}^f_k(t) = \check{C}^f_k(t)^*
    &= \frac{1}{2}\bra{\bar{0}}^*[e^{(i)^*H_{\text{SSH}}^*t}(\sY_0)^* e^{(-i)^*H_{\text{SSH}}^*t},\sum_j \cos(kj)(\sX_j)^*]_+\ket{\bar{0}}^*\\
    &=\frac{1}{2}\bra{\bar{0}}[e^{-iH_{\text{SSH}}t}(-\sY_0) e^{iH_{\text{SSH}}t},\sum_j \cos(kj)\sX_j]_+\ket{\bar{0}}\\
    &=-\check{C}^f_k(-t).
\end{aligned}
\end{equation}
Therefore, we can double the signal $\check{C}^f_k(t)$ in the negative time direction.

\section{Hardware implementation and calibration details}\label{app:Hardware calibration details}

\begin{figure}
    \centering
    \includegraphics[width=0.40\textwidth]{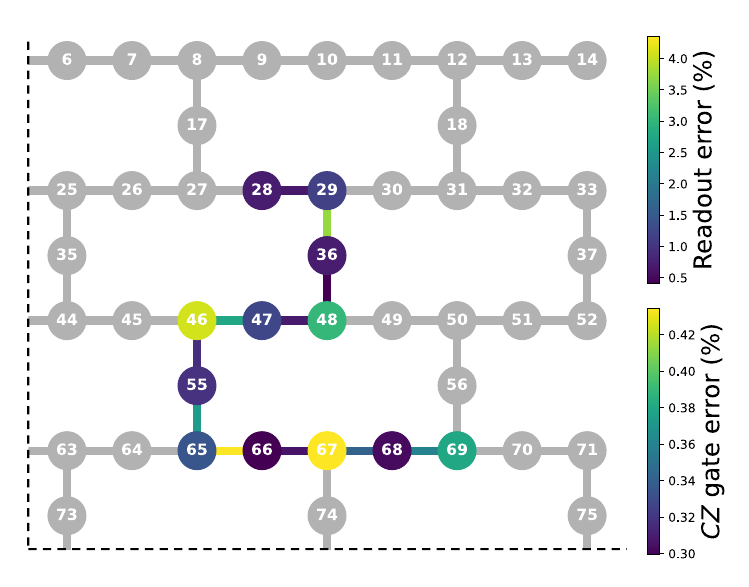}
    \cprotect\caption{(\textbf{a}) The 12 qubits of \verb|ibm_torino| used in the hardware implementation of the measurement circuits and their coupling map. The coupling map of \verb|ibm_torino| is colored to represent the readout error for each qubit and the two-qubit $CZ$ gate error for each qubit connection.}
    \label{fig:graph-circuit}
\end{figure}

In Table~\ref{tab:hardware-implementation-info}, we present details of the hardware implementation performed on IBM quantum chip \verb|ibm_torino|. During the execution of all quantum circuits, we use dynamic decoupling~\cite{Ezzel2023} and randomized compiling~\cite{PhysRevX.11.041039} to suppress noise in the quantum circuits. The measurements in the Schwinger model, SSH model, and TIM use numbers of qubits $12,12$ and $8$, respectively. In the hardware implementation, the 12 qubits are mapped to the physical qubits $(28,29,36,48,47,46,55,65,66,67,68,69)$ of \verb|ibm_torino| with their coupling map shown in Fig.~\ref{fig:graph-circuit}, and the 8 qubits are mapped to the physical qubits $(28,29,36,48,47,46,55,65)$. The coupling map is colored to represent the readout error rate for each qubit and the two-qubit $CZ$ gate error rate for each qubit connection. Other single-qubit properties of the 12 qubits are summarized in Table~\ref{table:qubit-properties}. All hardware data are obtained from the IBM cloud quantum platform~\cite{Qiskit} and more details are available in~\cite{Wang2025_correlation}.

\begin{table}[h]
\begin{tabular}{|c||c|c|c|c|c|c|c|c|c|}
\hline
\begin{tabular}[c]{@{}c@{}}Model\\ name\end{tabular}     & \begin{tabular}[c]{@{}c@{}}Hamiltonian\\ (parameters)\end{tabular}                          & \begin{tabular}[c]{@{}c@{}}$\tau$ \\ values\end{tabular} & \begin{tabular}[c]{@{}c@{}}Trotter \\ order\end{tabular} & \begin{tabular}[c]{@{}c@{}}Trotter \\ evolution \\ time\end{tabular} & \begin{tabular}[c]{@{}c@{}}Trotter  \\ step \\ length\end{tabular} & \# of $O_0$ & \# of circuits & \begin{tabular}[c]{@{}c@{}}\# of twirls \\ (per circuit)\end{tabular} & \begin{tabular}[c]{@{}c@{}}\# of shots \\ (per twirl)\end{tabular} \\ \hline\hline

Schwinger & \begin{tabular}[c]{@{}c@{}}$H_{\text{SW}}^{\text{tr}}$\\ $(m=0.5,g=0.3)$\end{tabular}       & 0.1                                                      & 2                                                        & 0-40                                                                 & 2                                                                  & 1           & 40             & 160                                                                   & 400                                                                \\ \hline
TIM       & $H_{\text{TIM}}$                                                                            & 0.1                                                   & 1                                                        & 0-11.8                                                               & 0.2                                                                & 1           & $60\times 2$   & 160                                                                   & 400                                                                \\ \hline
SSH       & \begin{tabular}[c]{@{}c@{}}$H_{\text{SSH}}$\\ $(v=1,\delta=0.8,$\\ $\mu=-2.5)$\end{tabular} & 0.1                                                      & 1                                                        & 0-23.6                                                               & 0.4                                                                & 7           & $60\times 7$   & 160                                                                   & 100                                                                \\ \hline
\end{tabular}
\cprotect\caption{Details of the measurement circuits in the Schwinger model, the transverse-field Ising model (TIM), and the Su-Schrieffer-Heeger (SSH) model. For each model, the second column gives the model parameters used in the hardware demonstration. The third column gives the $\tau$ values taken to measure the correlation functions. The fourth to sixth columns give the Trotter information in the quantum simulation of the Hamiltonian. The eighth column is the total number of quantum circuits we submitted to \verb|ibm_torino|. For all circuits, we use randomized compiling to mitigate errors led by coherent noise. The ninth and tenth columns record the number of Pauli twirls per circuit and the number of shots for each twirl in our implementation of randomized compiling.}
\label{tab:hardware-implementation-info}
\end{table}

\begin{table}[]
    \centering
\begin{tabular}{c|cccc}
\toprule
 & median & mean & min & max \\
\hline\hline
$\sqrt{X}$ error (\%)& 0.026 & 0.032$\pm$0.016 & 0.016 & 0.075 \\
Readout error (\%)& 1.21 & 1.79$\pm$1.40 & 0.42 & 4.35 \\
$T_1 (\mu s)$ & 209.71 & 185.24$\pm$52.84 & 62.73 & 232.57 \\
$T_2 (\mu s)$ & 138.79 & 135.72$\pm$66.73 & 47.12 & 292.97 \\
$CZ$ error (\%)& 0.342 & 0.349$\pm$0.046 & 0.300 & 0.435 \\
\hline
\end{tabular}
    \cprotect\caption{Summary of the single-qubit properties of the 12 physical qubits (28, 29, 36, 48, 47, 46, 55, 65, 66, 67, 68, 69) and their two-qubit $CZ$ error rates in \verb|ibm_torino|.}
    \label{table:qubit-properties}
\end{table}

\end{document}